\documentclass[
prd,preprint,nofootinbib,preprintnumbers,
longbibliography,
superscriptaddress,tightenlines
]{revtex4-1}

\usepackage{amsmath}    
\usepackage{graphicx}   
\usepackage{subfigure}
\usepackage{color}
\usepackage[colorlinks=true,linkcolor=blue,citecolor=blue, urlcolor=blue]{hyperref}   
\usepackage{bm} 
\usepackage{ulem}
\usepackage{soul}
\usepackage{graphicx}
\usepackage{amsmath,amssymb}
\usepackage{soul}
\usepackage{textcomp}
\usepackage{hyperref}
\usepackage{subfigure}
\usepackage[toc,page]{appendix}

\graphicspath{{./plot/}}

\def \be {\begin{equation} }
\def \ee {\end{equation}}
\def \bes {\begin{subequations} }
\def \ees {\end{subequations}}

\def \e {\epsilon}

\def \k {\kappa}
\def \o {\omega}

\def \s {\sigma}

\def \G {\Gamma}

\def \<{\langle}
\def \>{\rangle}
\def \+{\dagger}
\def \({\left(}
\def \){\right)}

\def \[{\left[}
\def \]{\right]}

\def\KZ{{\rm KZ}}

\def \mic {\rm{mic}}

\begin{document}

\title{The QCD critical point hunt: emergent new ideas and new dynamics}


\author{Yi Yin}
\email[]{yiyin3@mit.edu}
\affiliation{Center for Theoretical Physics, Massachusetts Institute of Technology, Cambridge, Massachusetts 02139, USA }

\preprint{MIT-CTP/5083}

\begin{abstract}

The QCD critical point is a landmark point on the QCD phase diagram, with potential connections to a plethora of deep questions on the properties of thermal nuclear matter. 
Future heavy-ion collision experiments, in particular the second phase of beam energy scan (BES-II), will explore the QCD phase diagram with an unprecedented precision and would potentially discover the QCD critical point. 
In this short review, I discuss and summarize recent new ideas, such as ``rapidity scan'', and new theoretical developments, in particular in the qualitative characterization and quantitative descriptions of the critical fluctuations in the expanding fireball. 
Those new ideas and developments together would enhance the prospect of locating the QCD critical point. 

%
\end{abstract}

\maketitle

\section{Introduction}
\label{sec:intro}

The past decade has seen significant advances on the characterization of the properties of thermal QCD matter at small baryon density, thanks to concerted experimental and theoretical efforts~\cite{Shuryak:2014zxa,Akiba:2015jwa}. 
The comparison between dynamical modeling predictions and experimental data have led to stringent constraint on transport properties of liquid quark-gluon plasma (QGP)~\cite{Heinz:2013wva,Heinz:2015tua}.  
At the same time, first principle lattice calculations have firmly established that the transition from baryon-neutral QGP to hadron gas is a rapid but smooth crossover~\cite{Ding:2015ona} ; 
the chiral pseudo-critical temperature $T_{c}$ is reported to be $156.5$~MeV with a remarkably small errorbar of $1.5$~MeV in a recent publication.~\cite{Steinbrecher:2018phh} (see also Ref.~\cite{DElia:2018fjp} for a brief review of recent progress).
In stark contrast to such quantitative precision,
our answers to some of the most important qualitative questions about QCD matter at finite baryon density (i.e. $\mu_{B}\geq 200$~MeV) are still speculative.  
Would QGP become more liquid-like or gas-like as we increase baryon density?
Would the phase transition separating the hadron gas and QGP become stronger and eventually turn out to be a first order transition? 
Are there any new phases? 
As explorers will sail from safe landscapes to look for new continents, as botanists will travel around the world to collect unknown species of plants, 
an important future direction of the field is exploring the baryon-rich regime of the QCD phase diagram~\cite{Geesaman:2015fha,Busza:2018rrf}. 
Indeed, the planed experimental programs in facilities worldwide, such as the second phase of the Beam Energy Scan Program (BES-II) at RHIC (U.S.), 
Compressed Baryonic Matter Experiment (CBM) at FAIR (Germany)~\cite{Ablyazimov:2017guv},  
``AFTER'' at LHC (Switzerland)~\cite{Hadjidakis:2018ifr}, ``CEE'' at HIAF (China), 
would open new observational frontier for the exploration of the QCD phase diagram in the next decades. 

One potential discovery of this exploration would be locating the conjectured QCD critical point, which is the end point of a line of first-order phase transitions~\cite{Stephanov:2007fk,Fukushima:2010bq}. 
The BES-II, starting next year (2019), will scan the phase diagram where we already saw interesting hints of criticality, and will bring data with unprecedented precision, 
see Refs.~\cite{Luo:2017faz} for a review. 
This unique opportunity in turn calls for timely theoretical efforts to maximize the discovery potential of this upcoming experiment.
The focus of this review will be on the recent theoretical developments in the context of the critical point search.

The critical point hunt will hinge on understanding and making full use of the critical fluctuations which grow universally near a critical point.  
This review will be organized by the quest for answering the following questions:
What new observables should we look into to enhance the prospect of discovering the critical point? 
What would be the qualitative features of the evolution of the critical fluctuations as the fireball expands, and how to quantitatively describe such evolution.
We shall see, for example, the emergence of new ideas on searching for the critical point by taking full advantage of the detector upgrades at STAR. 
We shall see emergent new scales which characterize the offequilibrium evolution of critical fluctuations, 
and new theoretical frameworks which could quantitatively describe the evolution of fluctuations themselves as well as their influence on the bulk evolution. 
Unlike many reviews of this kind, 
I will not talk about a story with a closed ending.
Instead, I would like to report developing news on the community's ongoing efforts towards building a new paradigm for quantitative study of the critical point.

\section{Enhancing the prospect of discovering the critical point: new ideas
}
\label{sec:exp}

The multiplicity fluctuations of the hadrons produced in heavy-ion collisions are sensitive to the presence of a critical point, where the conserved densities such as energy density and baryon density grow universally~\cite{Luo:2017faz}. 
By tuning the colliding energies $\sqrt{s}$ of two nuclei, experimentalists vary the baryon chemical potential $\mu_{B}$ (and temperature) of the nuclear matter created in heavy-ion collisions. 
Therefore, 
the nonmonotonic behavior of multiplicity fluctuations as a function of beam energy $\sqrt{s}$ have long been considered as promising signatures of the presence of the QCD critical point.


\begin{figure*}[!hbt]
\subfigure[\;]{\includegraphics[width=0.32\textwidth]{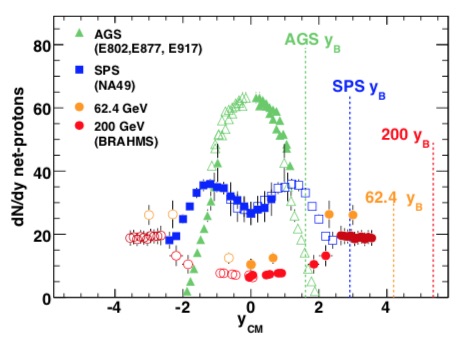}
\label{fig:NvsY}
}
\subfigure[\;]{\includegraphics[width=0.22\textwidth]{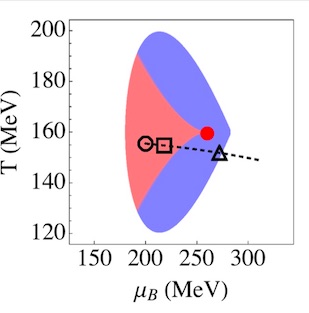} 
\label{fig:K4}
}
\subfigure[\;]{\includegraphics[width=0.36\textwidth]{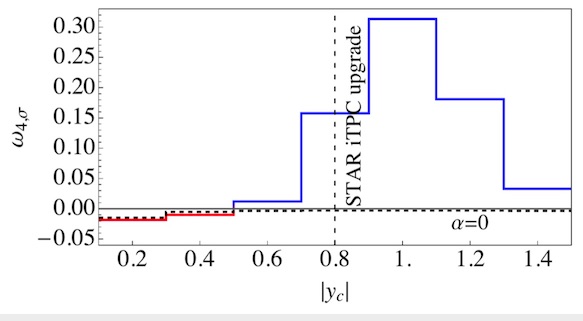}
\label{fig:o4}
}
\caption{
\label{fig:rapidity}
(color online)
%
The demonstration: the search for the QCD critical point via ``the rapidity scan''.
(a):
Net-proton distributions vs rapidity at different beam energies, indicating that baryon chemical potential $\mu_{B}$ depends on spacetime rapidity. 
The figure is taken from Ref.~\cite{Videbaek:2009zy}
(b):
The location of a hypothetical critical point (red circle) as well as changes in the freezeout conditions with increasing spacetime rapidity $y_{s}$, with the circle, square, and triangle indicating freezeout at $y_{s}=0,0.6, 1.2$ respectively. 
The colors denote the sign of the fourth cumulant of the order parameter field $K_{4}$, with $K_{4}<0$ in red and $K_{4}>0$ in blue.
(c): 
$\o_{4,\s}$, the rescaled proton fourth cumulants induced by critical fluctuations, computed in a pair of bins with width $\Delta y=0.4$ centered at $\pm y_{c}$ vs $y_{c}$. 
Note that for background fluctuations, the corresponding $\o_{4,\sigma}$ is of order $1$. 
The black dotted lines show results that are obtained by assuming boost-invariance. 
Such nonmonontonic dependence of the cumulants on rapidity can be employed to enhance the prospect of discovering the QCD critical point. 
\label{fig:rapidity}
}
\end{figure*} 

In fact, there are other knobs experimentalists could dial to vary the baryon chemical potential.
The thermal-fit of experimental data~\cite{Biedron:2006vf,Becattini:2007ci,Begun:2018efg} and theoretical model studies~\cite{Li:2016wzh,Shen:2018pty} indicate that
the baryon density also depends on spatial rapidity $y_{s}$ at fixed beam energy, and such dependence typically becomes stronger at a lower beam energy.
Complementary to scanning the phase diagram by taking steps in beam energy, 
examining the rapidity-dependence provides additional scans of small regions of the phase diagram. 
This idea of ``rapidity scan'' (RS) now emerges as a new method, and has already attracted much experimental and theoretical interest. 
For example, a proposal for a possible future fixed target program at the LHC is now exploring the scientific opportunities brought by RS~\cite{Hadjidakis:2018ifr}. 
 
We now focus on the search for the QCD critical point based on RS.
Brewer, Mukherjee, Rajagopal and myself propose to bin the cumulants in rapidity to give a more crisp picture of the critical regime~\cite{Brewer:2018abr}. 
To demonstrate this idea in a simplified set-up,
we consider a hypothetical scenario that at midrapidity, the freezeout happens in a region where the fourth cumulant of order parameter field $K_{4}$ is negative (see black circle in Fig.~\ref{fig:K4}). 
However, as we increase $y_{s}$, the chemical potential becomes larger, and the corresponding $K_{4}$ would change sign as implied by the critical universality. 
Indeed,  the sign change of the fourth cumulants $K_{4}$ with increasing beam energy was predicted in Ref.~\cite{Stephanov:2008qz}. 
In our work instead, we consider the dependence of $\o_{4,\s}\equiv \k_{4,N}/N_p$ on the center of the bin $y_{c}$ where $\kappa_{4,N}$ denotes the fourth cumulants of proton numbers induced by critical fluctuations and $N_{p}$  denotes the number of protons in the rapidity bin. 
Observe as expected that  $\o_{4,\s}$ is negative for small $y_{\max}$ but eventually become positive at a larger $y_{\max}$ as Fig.~\ref{fig:o4} illustrates. 
It is worth noting that the success of the ``rapidity scan'' would crucially rely on  a large rapidity coverage as well as on excellent particle identification capabilities. 

I wish to close this section by briefly mentioning other new ideas proposed in a previous couple of years related to the critical point search. 
 It would be interesting to explore HBT-like observable~\cite{Plumberg:2017tvu} as the enhanced critical fluctuations are tied to enhanced correlations. 
To isolate fluctuations of baryon density distribution in initial state~\cite{Bzdak:2016jxo}, one might also try a creative way of binning the events, see Ref.~\cite{Akamatsu:2018vjr} for a preliminary proposal.
In addition to the universal aspect of the QCD critical point, 
it is also useful to take the peculiarity of the theory of QCD into consideration. 
Studies along this line of thought include the discussion of the observational effects due to the modification of nuclear force~\cite{Shuryak:2018lgd} as well as the modificaition of light nuclei production~\cite{Sun:2018jhg} near the critical point.

\section{Towards a quantitative framework to the critical point search}

\subsection{Why hydrodynamics is not enough}

To turn high precision experimental data anticipated in the upcoming years, including the measurement of the new observables mentioned in the previous section,  
into definite information on the QCD critical point, 
a quantitative framework for modeling the salient features of these low energy collisions would be indispensable. 
Viscous hydrodynamic modeling is a key ingredient in the state-of-art description of the evolution of the fireballs created at top RHIC and LHC energies. 
Progress has been made recently on extending this successful program to lower collision energies~\cite{Noronha:2018atu}.
This includes the study of the initial conditions at BES energies~\cite{Shen:2017bsr,Okai:2017ofp,Akamatsu:2018olk}, 
and subsequent hydrodynamic evolution~\cite{Denicol:2018wdp,Du:2018mpf},
see also Refs~\cite{Petersen:2017jdb,Shen:2018pty} for more references.
If a fireball is passing close to the critical point, 
pertinent features associated with a critical point has to be included as well. 
For example, 
a family of equation of state (E.o.S)  with a critical point in the 3D Ising universality class has been constructed in Ref.~\cite{Parotto:2018pwx}, which significantly improves the early work~\cite{Nonaka:2004pg}.
Each E.o.S matches to available lattice results in the small $\mu_{B}$ region, 
but is different from others due to the location of the critical point as well as other parameters which controls the mapping between the QCD critical point and the 3D Ising model.
And, the parametrization of transport coefficients with appropriate critical behavior is also needed~\cite{Kapusta:2012zb,Monnai:2016kud}.
Since the parametrization of the E.o.S and transport coefficients depends on the location of the critical point, 
a comparison between the data and the output of a reliable quantitative modeling would eventually constrain the parameter space for the critical point in the future.

However, as the fireball approaches the critical point, the paradigm based on hydrodynamic modeling is not enough. 
In particular, if nature is generous enough to put a critical point in the regime accessible to heavy-ion collision experiments, 
nature would also hide her secret by offequilibrium effects which would happen inescapably near the critical point, as we shall see in the next subsection.

\subsection{The characteristic features of offequilibrium effects near the critical point
\label{sec:KZ}
}

\begin{figure*}
\begin{center}
\subfigure[\;]{\includegraphics[width=0.32\textwidth]{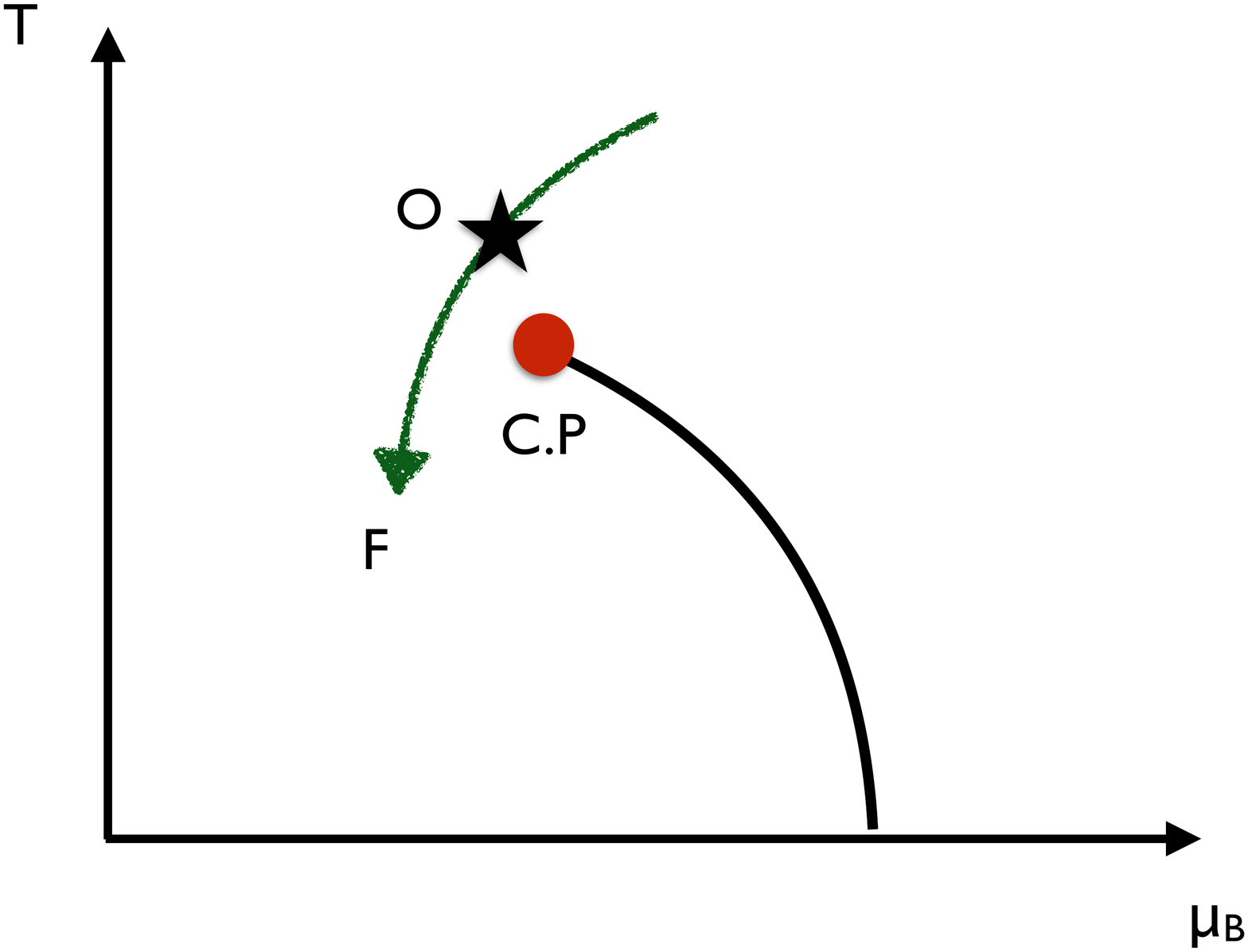}
\label{fig:traj}
}
\subfigure[\;]{\includegraphics[width=0.32\textwidth]{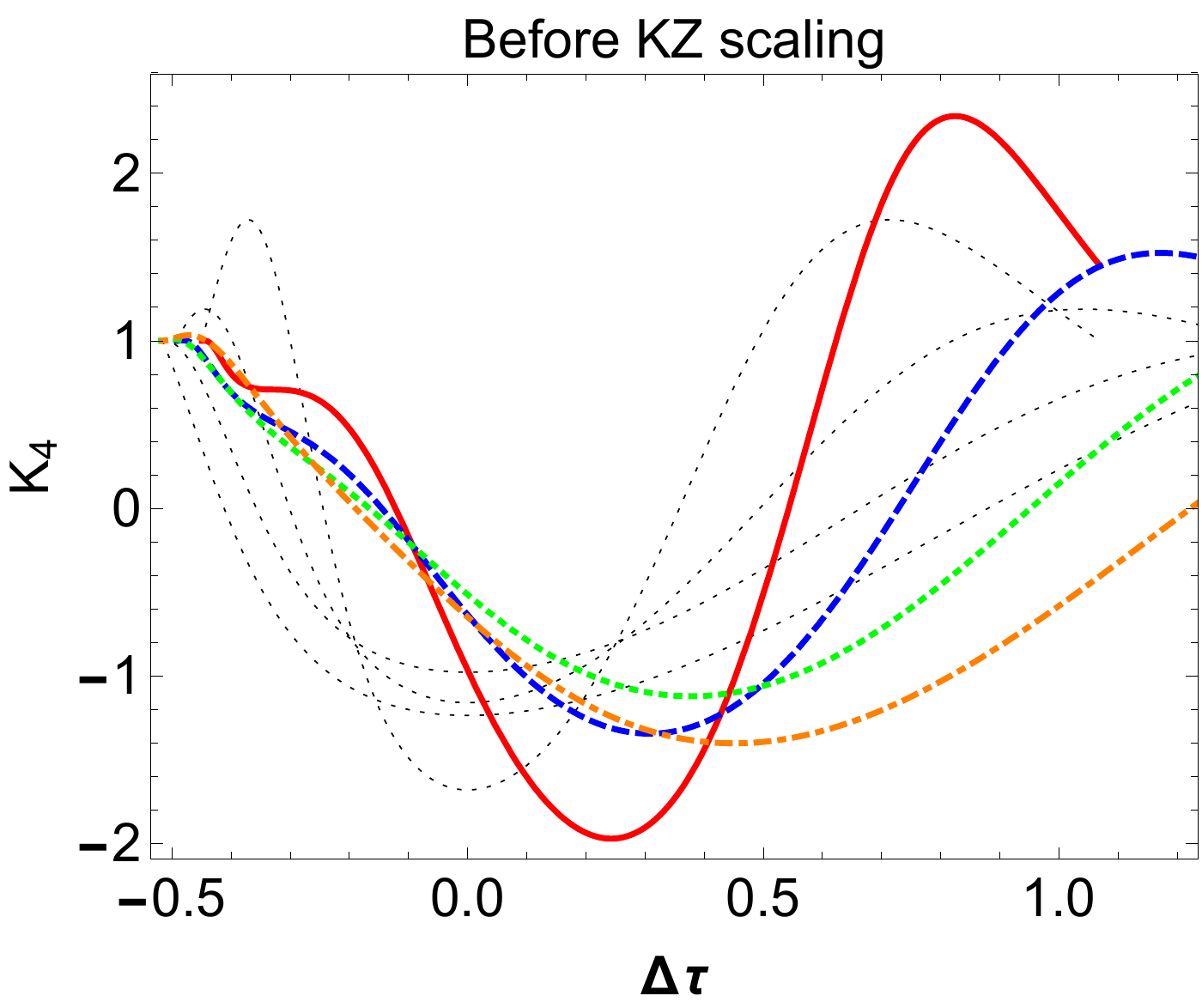}
\label{fig:K4raw} 
}
\subfigure[\;]{\includegraphics[width=0.32\textwidth]{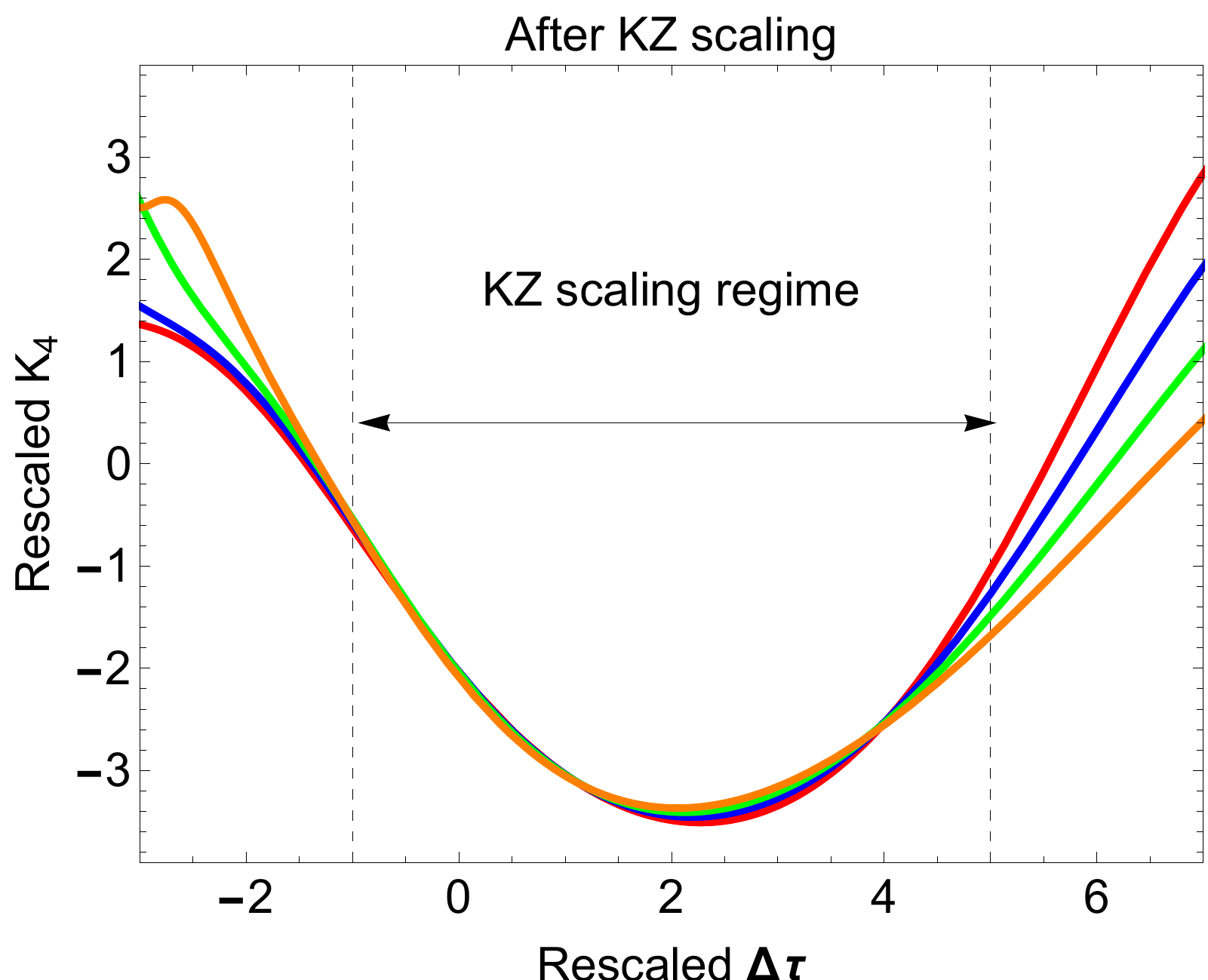}
\label{fig:K4scale}
}
\caption{
(color online) 
An illustration of the emergence of the offequilibrium effect and of offequilibrium scaling with respect to the Kibble-Zurek length $l_{\rm KZ}$ and time $\tau_{KZ}$ (see text)
(a):
A schematic trajectory (green solid curve) of a fireball created in a heavy-ion collision passing close to the critical point. 
The point ``F'' illustrates where the freezeout happens. 
Solid black curve illustrates the first order transition line with the end point (red circle) corresponding to the critical point (C.P.). 
The fireball would fall out of equilibrium before reaching the critical point at some point on the trajectory as illustrated by point ``O''.
(b):
Temporal evolution of the kurtosis $K_{4}$ (normalized
by its initial equilibrium value) for four different representative trajectories. 
 Their corresponding equilibrium values are plotted in dotted curves. 
 $\Delta \tau$ denotes the proper time of the evolution and becomes zero by construction whenever the trajectory is the closest to the critical point. 
 (c): 
 The rescaled $K_{4}$ with respect to KZ length $l_{KZ}$ vs the rescaled time $\Delta\tau/\tau_{\rm KZ}$, see text and Ref.~\cite{Mukherjee:2016kyu} for more details. 
}  
\end{center}
\vspace{-0.2in}
\end{figure*}

To see why offequilibrium effects become unavoidable near the critical point, let us recall that thermodynamic fluctuations will be equilibrated through a diffusive process. 
The equilibration rate $\G(Q)$ at a given momentum $Q$ behaves as
\footnote{Strictly speaking, the $Q$ dependence of $\G(Q)$ near a critical point is more complicated than Eq.~\eqref{diffusion} due to the non-linear effects~\cite{onuki2002phase}.
However, such details are not important for the qualitative discussion presented in this paragraph.} 
\begin{eqnarray}
\label{diffusion}
\G(Q) = D\, Q^{2}\, , 
\end{eqnarray}
where $D$ denotes the diffusive constant. 
The fluctuation-dissipation relation further tells us 
\begin{eqnarray}
\label{eq:fluct-diss}
D
\propto 
\frac{\textrm{transport coefficient}}{\textrm{the magnitude of fluctuations}}  \, .
\end{eqnarray}
Therefore whenever the fireball approaches the critical point, 
the critical fluctuation would grow and the equilibration rate would become small (the phenomena of the ``critical slowing down''). 
Eventually, the fireball would fail to catch up with the growth of equilibrium fluctuations at some turning point (denoted by ``O'' in Fig.~\ref{fig:traj} ) along its trajectory in the phase diagram,
and it would afterward fall out of equilibrium. 
As a result, 
the fireball could only ``memorize'' information along its evolution up to the point ``O''  at best, 
in analog to ``jet-lag'' that one would feel when crossing multiple time zones very quickly.

We now discuss an interesting and phenomenologically important example, which demonstrates that the ``jet-lag'' effect would even change the qualitative feature of critical fluctuations.
Here, we focus on $K_{3}$, 
the third cumulant of the order parameter field.
Its equilibrium value is negative above the crossover line and positive below the crossover line, 
where freezeout (see point ``F'' in Fig.~\ref{fig:traj}) occurs. 
Then, if critical fluctuations were in equilibrium, 
$K_{3}$ is expected to make the skewness of multiplicity fluctuations of protons larger than the baseline value.
Turning to the preliminary experimental result reported by STAR~\cite{Luo:2015doi}, 
the data is below the baseline, and hence is opposite to the equilibrium expectation (see also Ref.~\cite{Jiang:2015hri}). 
This proton skewness ``sign puzzle'' can be naturally explained by noting the effect of  ``jet-lag''~\cite{Mukherjee:2015swa},
as the fluctuations that the fireball could memorize would be ``frozen'' at point ``O'' in Fig.~\ref{fig:traj}, 
where the value of $K_{3}$ memorized by the fireball would likely be negative.
Indeed, it has been demonstrated in Ref.~\cite{Mukherjee:2015swa} as well as in many other studies~\cite{Nahrgang:2018afz} that 
both Gaussian and non-Gaussian fluctuations can be quantitatively and even qualitatively different from equilibrium expectations.

While the equilibrium critical scaling is no longer applicable to the expanding fireball, 
the qualitative features of offequilibrium evolution near the critical point might be captured by employing the key idea of Kibble-Zurek (KZ) dynamics. 
This idea was pioneered by Kibble in a cosmological setting, and was generalized to describe a similar problem in condensed matter system (see Ref.~\cite{Zurek:1996sj} for a review). 
The key observation is that since the evolution of the critical fluctuations become effectively frozen at the point (i.e. the point ``O'' in Fig.~\ref{fig:traj}) where the time remaining to reach the critical point is shorter than the relaxation time, 
then why not measure those fluctuations and their evolutions in the units of KZ length $l_{\KZ}$ and KZ time $\tau_{\KZ}$ where
they are the correlation length and the relaxation time at point ``O'' respectively.

In Ref.~\cite{Mukherjee:2016kyu}, an extension of KZ dynamics to situations relevant to the search for the QCD critical point has been made by Mukherjee, Venugopalan and myself. 
As an example, 
we consider four different trajectories passing close to the critical point with different expansion rates. 
In Fig.~\ref{fig:K4raw}, we see that the magnitude as well as the temporal dependence of $K_{4}$ along those trajectories are different from each other. 
We then do the rescaling, and present $K_{4}$ rescaled by an appropriate power of $l_{\KZ}$ vs rescaled proper time $\tau/\tau_{\KZ}$ in Fig.~\ref{fig:K4scale}
While the corresponding $\tau_{\KZ}$ and $l_{\KZ}$ are different for each trajectory, 
the rescaled kurtosis curves beautifully collapse into a single curve in the vicinity of the critical point. 
This is just one example which demonstrates the emergent scales, $l_{\KZ}$ and $\tau_{\KZ}$, characterizing the magnitude and temporal evolution of the critical fluctuations respectively, as discussed in Ref.~\cite{Mukherjee:2016kyu}. 
Such scaling properties open new possibilities to identify experimental signatures of the critical point.
Recently, KZ offequilibrium scaling has been explored and verified in a variety of materials, such as hexagonal manganites~\cite{PhysRevX.2.041022} and optical lattice~\cite{Clark606}. 
It would be interesting to see if nuclear matter would exhibit such offequilibrium scaling in future BES or ``rapidity scan'', though background effects should be subtracted with care.

Let us further appreciate the rich physics behind $l_{\KZ}$ and $\tau_{KZ}$ from a different but related perspective. 
Since the equilibration rate $\G(Q)$ in Eq.~\ref{diffusion} becomes smaller with smaller $Q$,
there must exist a characteristic wavelength above which the fluctuations fall out of equilibrium for a system passing close to the critical point. 
It turns out that this length scale is nothing but $l_{\KZ}$, and the critical fluctuations become the strongest at scale $l_{\KZ}$~\cite{Akamatsu:2018vjr}. 
This implies that critical fluctuations at length scale $l_{KZ}$, which will be referred to as ``KZ modes'' from now on, 
contain the most information about the criticality.
To add,
the lifetime of ``KZ modes'' is of order $\tau_{\KZ}$ as one can easily verify. 
In fact, the above is the underlying physics that $l_{\KZ}$ characterizes the magnitude of off-equilibrium evolution and $\tau_{\KZ}$ characterizes their temporal evolution.

We now have the following picture to ``visualize'' the essential physics for a fireball passing close to the critical point. 
Before the fireball enters the critical regime, 
the fireball can be divided into several ``droplets'', and each droplet can be considered to be in local equilibrium. 
The sizes of such droplets are of the order $l_{\max}$, 
the maximum wavelength that can be reached by diffusion,
and the fluctuations at scale shorter than $l_{\max}$ are in equilibrium. 
As the fireball approaches the critical point, 
the critical fluctuations at typical scale $l_{\KZ}$, i.e. ``KZ modes'', become offequilibrium
Since the diffusive constant $D$ is highly suppressed near the critical point, 
there is an interesting hierarchy among $l_{\max}$, $l_{\KZ}$ and the microscopic length $l_{\mic}$~\cite{Akamatsu:2018vjr}:
\begin{eqnarray}
\label{hierachy}
l_{\mic}\ll l_{\rm KZ}\ll l_{\rm{max}}\, . 
\end{eqnarray}
whenever the microscopic relaxation time is much shorter than the expansion time of the fireball, which is of the order $10$~fm. 
This hierarchy means that ``KZ modes'' are otherwise equilibrated macroscopic modes that fall offequilibrium near the critical point. 
Therefore they may be considered as emergent new degrees of freedom near the critical point with a macroscopic lifetime $\tau_{\KZ}$.

One of the main themes of this section is to emphasize the importance of offequilibrium effects as well as the emergence of KZ scales. 
Historically, the universal equilibrium critical behavior provided very useful guidance for the critical point search. 
However,  its domain of validity ends before, and in fact well before, the regime of interest where critical fluctuations are significantly enhanced, see Ref.~\cite{Akamatsu:2018vjr} for a detailed analysis. 
This observation makes clear that simply inputting an E.o.S (and transport coefficients) with (equilibrium) scaling behavior into conventional hydrodynamic simulation would not be enough to describe the bulk evolution. 
In such offequilibrium regime, a new set of degrees of freedom, namely ``KZ modes'',  emerge.
It is imperative to describe those emergent degrees of freedom for the purpose of quantitative study near the critical point, 
and this will be addressed in the subsequent section.

%

\subsection{Dynamical framework: from hydrodynamics to fluctuating hydrodynamics}
\label{sec:framework}

Let us begin by briefly reviewing the previous studies associated with dynamics of critical fluctuations.
The limitation of the growth of the critical correlation length due to finite time effects was studied in Ref.~\cite{Berdnikov:1999ph}.
A closed-form expressions for the temporal evolution of both Gaussian and non-Gaussian cumulants of order parameter fields are derived in Ref.~\cite{Mukherjee:2015swa} and are applied to study offequilibrium effects on critical cumulants.
Meanwhile,
the model of chiral fluid dynamics (CFD) has been developed over recent years~\cite{Herold:2013bi,Herold:2014zoa,Herold:2016uvv} (see Ref.~\cite{Nahrgang:2016ayr} for more references).
In the CFD framework, the chiral condensate is identified as a dynamical variable.
However, as already pointed in Refs.~\cite{Son:2004iv,Fujii:2004jt} (see also Ref.~\cite{Fukushima:2010bq}), 
the chiral condensate would be expressible as a function of energy density and baryon density at a macroscopic scale,
and should not be counted as an additional slow degree of freedom. 
Therefore the appropriate quantitative framework would be \textit{fluctuating hydrodynamics},  which describes the dynamics of (average) hydrodynamic variables and their fluctuations, 
with the implementation of the salient feature of a critical point. 
Below, 
I will first discuss recent developments in fluctuating hydrodynamics in general before moving to the special case of critical fluctuating hydrodynamics.

The conventional approach to fluctuating hydrodynamics treats hydrodynamic variables as stochastic variables, and will be referred as the ``stochastic approach'' hereafter.
Stochastic noise is added to the hydrodynamic equations with magnitudes fixed by the fluctuation-dissipation theorem, as was done in Ref.~\cite{landau2013fluid} for non-relativistic fluid. 
This approach has been extended to relativistic hydrodynamics in Ref.~\cite{Kapusta:2011gt}, 
see Refs~\cite{Murase:2013tma} for related developments. 
Even though numerical simulations based on ``stochastic approach'' are computationally demanding, 
encouraging new results are reported in Ref.~\cite{Singh:2018dpk} where stochastic noise is implemented in a 3D hydrodynamic simulation for baryon-neutral QGP.
See also Refs.~\cite{Kapusta:2012zb,Sakaida:2017rtj} for related studies near the critical point.
%
%

In parallel, 
there is growing interest, mostly from the high energy physics community, in constructing or deriving  the action corresponding to fluctuating hydrodynamics~\cite{lebedev1983diagram,Grozdanov:2013dba, Kovtun:2014hpa,Crossley:2015evo,Glorioso:2017fpd}.
I will refer this approach as the``action approach''.
While direct application of the ``action approach'' to heavy-ion collision phenomenology would not be possible at this moment, 
such an approach has already provided tanalytic insight into some salient features of fluctuating hydrodynamics.

Very recently, a third approach, namely the ``hydro+ approach''~\cite{Stephanov:2017ghc} or ``hydro-kinetic approach''~\cite{Akamatsu:2016llw} has emerged,
though early studies in a similar fashion can be traced back to 1970's~\cite{KAWASAKI19701,andreev1978corrections}.
The key idea is to directly consider wavenumber-dependent correlation functions of hydrodynamic variables as slow variables in addition to hydrodynamic ones.  
The resulting equations of motion are deterministic, and describe the coupled evolution among those correlation functions and conventional hydrodynamic variables.  
This approach successfully describes non-trivial offequilibrium effect such as the hydrodynamic tail~\cite{Akamatsu:2016llw} and the renormalization of bulk viscosity for a fluid away from~\cite{Akamatsu:2017rdu} or near a critical point~\cite{Stephanov:2017ghc}.

The formulation of fluctuating hydrodynamics in the vicinity of the QCD critical point using the third approach was first presented in Ref.~\cite{Stephanov:2017ghc} by Stephanov and myself, and the resulting theory is referred as ``hydro+''. 
Since fluctuation of the specific entropy density $\hat{s}$ (the ratio of entropy to density baryon) will always be proportional to the corresponding fluctuations of the order parameter field in the Ising model (see Refs~\cite{Stephanov:2017ghc,Akamatsu:2018vjr} for the derivation), 
we ``add'' the wavenumber-dependent two point function of $\hat{s}$ to the hydrodynamic equations. 
Following Ref.~\cite{Stephanov:2017ghc}, we will denote this two point function by $\phi$ here with its dependence on space-time and momentum suppressed. 
The equilibration of $\phi$ is described by a relaxation rate equation with equilibration rate $\Gamma(Q)$ known from the dynamical universality~\cite{onuki2002phase}.  
While the conservation equations such as $\partial_{\mu}\, T^{\mu\nu}=0$ are still in their usual forms
, 
$\phi$ is coupled to the bulk evolution 
as the local equilibrium pressure $p(\e,n)$ in the constitutive relation of hydrodynamics is replaced by a generalized pressure $p_{+}(\e,n,\phi)$ which now depends on $\e, n$ and the critical fluctuation $\phi$.
Remarkably,  this generalized pressure can be derived systematically~\cite{Stephanov:2017ghc}. 
Likewise, we also need to replace transport coefficients such as bulk viscosity with the generalized ones.

As an illustration, we compute the ``effective sound velocity'' $c^{2}_{s,,{\rm eff}}$ as a function of the expansion rate of a critical fluid by solving the linearized ``hydro+'' equations in Ref.~\cite{Stephanov:2017ghc}. 
When the expansion rate becomes much smaller than $1/\tau_{\KZ}$ (where $\tau_{\KZ}$ is the Kibble-Zurek introduced in Sec.~\ref{sec:KZ}), 
the critical fluctuation $\phi$ is in equilibrium and $c^{2}_{s,,{\rm eff}}$ approaches its equilibrium value. 
Critical fluctuations $\phi$ would fall out of equilibrium as expansion rate becomes larger than $1/\tau_{\KZ}$, 
and ``effective sound velocity'' becomes larger and larger. 
What we learned here is that applicability of ``vanilla'' hydrodynamic is limited at time scale much longer than $\tau_{\KZ}$.
Naively extending hydrodynamics to a shorter time scale would underestimate the stiffness of E.o.S. 
Instead, one has to include the additional slow evolution of $\phi$, most notably from ``KZ modes'', and this physics is captured in the framework of ``hydro+''.


The main message of this subsection is that there are rapid and significant developments, at both the technical and conceptual level, on the application of fluctuating hydrodynamics to the QCD critical point.
This marks the beginning of the quantitative era of studying of critical dynamics. 
Very recently, a simulation of baryon density fluctuations near a critical point based on the ``stochastic approach'' has been reported in Ref.~\cite{Nahrgang:2018afz} in a simplified geometry.
At the time of writing this review, 
a numerical simulation of ``hydro+'' in a simplified setting is near completion~\cite{Greg-Ryan}. 
The future directions of those quantitative study include adding needed sophistication to those modelings as well as the formulation of an appropriate freezeout prescription which turns critical fluctuations into hadron multiplicity fluctuations. 

To close this section, 
let me speculate on how the qualitative understanding presented in Sec.~\ref{sec:KZ} might be used to make progress in quantitative studies and to connect different approaches. 
We have discussed that there is a characteristic wavelength around which the dynamics of fluctuations are most important. 
This length scale is identified as $l_{\rm{max}}$ for a system away from the critical point in Ref.~\cite{Akamatsu:2016llw} and as the Kibble-Zurek length $l_{\rm{KZ}}$ in Ref.~\cite{Akamatsu:2018vjr}. 
Tracing modes with wavelength much shorter than this characteristic wavelength is unnecessary as they are equilibrated. 
Once those boring modes are ``thrown out'', the computation cost using the ``stochastic approach'' would be significantly reduced, see for example the study of Ref.~\cite{Singh:2018dpk}.
On the other hand, 
the emergence of the separation of length scale in Eq.~\eqref{hierachy} suggests that one could further coarse-grain the action obtained in the ``action approach'', following the standard  "Wilsonian picture". 
It would be interesting to see if this ``Wilsonian picture'' would lead to the equations used in the third approach, perhaps along the line of Ref.~\cite{Chris}.

\section{Summary and outlook}
\label{sec:conclusion}

In this review, I discuss and summarize recent new ideas and theoretical developments towards maximizing the discovery potential of the upcoming search for the QCD critical point. 
In particular, the idea of ``rapidity scan'' can be used as a new method complementary to the conventional beam energy scan to study the properties of baryon-rich QCD matter, including locating the critical point. 
The comprehensive quantitative dynamic framework to describe the evolution of baryon-rich fireballs is under rapid construction. 
The focus of this review is on the critical dynamics. 
As the fireball approaches the critical point, 
the competition between the expansion and equilibration induces the ``jet-lag'' effect which upsets naive interpretation of the data based on equilibrium expectation, and ``excite'' offequillibrium fluctuations 
with characteristic wavelength and lifetime determined by Kibble-Zurek (KZ) dynamics. 
Those emergent ``KZ modes'' would in turn strongly influence the expansion of the fireball, and such influence can be quantitatively described by the fluctuating hydrodynamics with a critical point using the ``stochastic approach'' and the ``hydro+'' (or ``hydro-kinetic'') approach.  
%
%
%
%
%

In a broad aspect, 
the competition between the expansion and equilibration is one of the common themes in many other studies of heavy-ion collision physics, 
in particular in the study of initial thermalization, of extracting transport coefficient from the ``memory'' of initial geometry, and of the small colliding systems.
It remains to see the cross-fertilization among those studies and the study of critical dynamics. 
To add,
the investigation of the fluctuating hydrodynamics has strong cross-disciplinary appeal, with applications in the condensed matter physics (e.g. Ref~\cite{PhysRevLett.114.230602}), and cosmology (e.g.Ref~\cite{Carrasco:2012cv,Floerchinger:2016hja}).

In this somewhat subjective review, many important directions/achievements are not covered.
In particular,
seeking for the signature associated the first order transition would not only be instrumental to locate the critical point, but also advance our understanding of the QCD phase structure. 
This structure would also be probed by the signatures pertinent to the chiral symmetry restoration.
Such signatures might be revealed through the spectrum of dilepton as well as observables sensitive to chiral effects which are macroscopic manifestation of the microscopic quantum anomaly~\cite{Kharzeev:2015znc}. 

Two decades ago, our knowledge of baryon-neutral QGP was also very poor. 
However the running of heavy-ion collision experiments at RHIC, and then at LHC, together with consistent theoretical efforts have changed all that. 
In the light of this, 
I await with great interest the results from both future experiments and the output of dynamical modeling.
And, I believe there is a great scientific opportunity awaiting us.  
I will take this opportunity to participate such future studies of the critical point search and of the exploration of the QCD phase
that would potentially bring the emergence of new insight and even the shift of paradigm. 
I hope this review would motivate the reader to do the same.

%
%
%

%
%

{\bf Acknowledgment:} 
I thank Yukinao Akamatsu, Jasmine Brewer, Chris Lau, Hong Liu, Akihiko Monnai, Swagato Mukherjee, Krishna Rajagopal, Gregory Ridgway, Misha Stephanov, Derek Teaney, Raju Venuogaplan, Ryan Weller, Fanglida~Yan for fruitful collaboration on a number of works discussed in this review. 
In addition, I am grateful to Marcus Blum, Lipei Du, Ulrich Heinz, Jiangyong Jia, Lijia Jiang, Che-Ming Ko, Roy Lacey, Xiaofeng~Luo, Marlene Nahrgang, Xu~Nu, Rob Pisarski, Scott Pratt, Chun~Shen, Thomas Schaffer, Bjorn Schenke, Mayank Singh, Huichao~Song for helpful conversation. 
I would like to thank Lipei Du, Jasmine Brewer, Xiaofeng Luo for insightful comments on the draft,
I thank the organizers of QM 2018 for the invitation to give a plenary talk on the critical point search that motivates me to start drafting this review, 
and thank Lijuan~Ruan for her kind message during the difficult time of preparing this plenary talk. 
Finally, I thank the theoretical physics department at CERN,  Quark Matter Research Center at Institute of Modern Physics (IMP) in China for hospitality, where part of this work has been done.  
The research is supported by 
the Office of Nuclear Physics of the U.S. Department of Energy under Contract Number DE-SC0011090 and within the framework of the Beam Energy Scan Theory (BEST) Topical Collaboration.


\bibliography{ref}

\begin{thebibliography}{75}%
\makeatletter
\providecommand \@ifxundefined [1]{%
 \@ifx{#1\undefined}
}%
\providecommand \@ifnum [1]{%
 \ifnum #1\expandafter \@firstoftwo
 \else \expandafter \@secondoftwo
 \fi
}%
\providecommand \@ifx [1]{%
 \ifx #1\expandafter \@firstoftwo
 \else \expandafter \@secondoftwo
 \fi
}%
\providecommand \natexlab [1]{#1}%
\providecommand \enquote  [1]{``#1''}%
\providecommand \bibnamefont  [1]{#1}%
\providecommand \bibfnamefont [1]{#1}%
\providecommand \citenamefont [1]{#1}%
\providecommand \href@noop [0]{\@secondoftwo}%
\providecommand \href [0]{\begingroup \@sanitize@url \@href}%
\providecommand \@href[1]{\@@startlink{#1}\@@href}%
\providecommand \@@href[1]{\endgroup#1\@@endlink}%
\providecommand \@sanitize@url [0]{\catcode `\\12\catcode `\$12\catcode
  `\&12\catcode `\#12\catcode `\^12\catcode `\_12\catcode `\%12\relax}%
\providecommand \@@startlink[1]{}%
\providecommand \@@endlink[0]{}%
\providecommand \url  [0]{\begingroup\@sanitize@url \@url }%
\providecommand \@url [1]{\endgroup\@href {#1}{\urlprefix }}%
\providecommand \urlprefix  [0]{URL }%
\providecommand \Eprint [0]{\href }%
\providecommand \doibase [0]{http://dx.doi.org/}%
\providecommand \selectlanguage [0]{\@gobble}%
\providecommand \bibinfo  [0]{\@secondoftwo}%
\providecommand \bibfield  [0]{\@secondoftwo}%
\providecommand \translation [1]{[#1]}%
\providecommand \BibitemOpen [0]{}%
\providecommand \bibitemStop [0]{}%
\providecommand \bibitemNoStop [0]{.\EOS\space}%
\providecommand \EOS [0]{\spacefactor3000\relax}%
\providecommand \BibitemShut  [1]{\csname bibitem#1\endcsname}%
\let\auto@bib@innerbib\@empty
\bibitem [{\citenamefont {Shuryak}(2017)}]{Shuryak:2014zxa}%
  \BibitemOpen
  \bibfield  {author} {\bibinfo {author} {\bibfnamefont {Edward}\ \bibnamefont
  {Shuryak}},\ }\bibfield  {title} {\enquote {\bibinfo {title} {{Strongly
  coupled quark-gluon plasma in heavy ion collisions}},}\ }\href {\doibase
  10.1103/RevModPhys.89.035001} {\bibfield  {journal} {\bibinfo  {journal}
  {Rev. Mod. Phys.}\ }\textbf {\bibinfo {volume} {89}},\ \bibinfo {pages}
  {035001} (\bibinfo {year} {2017})},\ \Eprint {http://arxiv.org/abs/1412.8393}
  {arXiv:1412.8393 [hep-ph]} \BibitemShut {NoStop}%
\bibitem [{\citenamefont {Akiba}\ \emph {et~al.}(2015)\citenamefont {Akiba}
  \emph {et~al.}}]{Akiba:2015jwa}%
  \BibitemOpen
  \bibfield  {author} {\bibinfo {author} {\bibfnamefont {Yasuyuki}\
  \bibnamefont {Akiba}} \emph {et~al.},\ }\bibfield  {title} {\enquote
  {\bibinfo {title} {{The Hot QCD White Paper: Exploring the Phases of QCD at
  RHIC and the LHC}},}\ }\href@noop {} {\  (\bibinfo {year} {2015})},\ \Eprint
  {http://arxiv.org/abs/1502.02730} {arXiv:1502.02730 [nucl-ex]} \BibitemShut
  {NoStop}%
\bibitem [{\citenamefont {Heinz}(2013)}]{Heinz:2013wva}%
  \BibitemOpen
  \bibfield  {author} {\bibinfo {author} {\bibfnamefont {Ulrich~W.}\
  \bibnamefont {Heinz}},\ }\bibfield  {title} {\enquote {\bibinfo {title}
  {{Towards the Little Bang Standard Model}},}\ }\bibfield  {booktitle} {\emph
  {\bibinfo {booktitle} {{Proceedings, Workshop on Discovery Physics at the LHC
  (Kruger 2012): Kruger National Park, Mpumalanga, South Africa, December 3-9,
  2012}}},\ }\href {\doibase 10.1088/1742-6596/455/1/012044} {\bibfield
  {journal} {\bibinfo  {journal} {J. Phys. Conf. Ser.}\ }\textbf {\bibinfo
  {volume} {455}},\ \bibinfo {pages} {012044} (\bibinfo {year} {2013})},\
  \Eprint {http://arxiv.org/abs/1304.3634} {arXiv:1304.3634 [nucl-th]}
  \BibitemShut {NoStop}%
\bibitem [{\citenamefont {Heinz}\ \emph {et~al.}(2015)\citenamefont {Heinz}
  \emph {et~al.}}]{Heinz:2015tua}%
  \BibitemOpen
  \bibfield  {author} {\bibinfo {author} {\bibfnamefont {Ulrich}\ \bibnamefont
  {Heinz}} \emph {et~al.},\ }\bibfield  {title} {\enquote {\bibinfo {title}
  {{Exploring the properties of the phases of QCD matter - research
  opportunities and priorities for the next decade}},}\ }\href@noop {} {\
  (\bibinfo {year} {2015})},\ \Eprint {http://arxiv.org/abs/1501.06477}
  {arXiv:1501.06477 [nucl-th]} \BibitemShut {NoStop}%
\bibitem [{\citenamefont {Ding}\ \emph {et~al.}(2015)\citenamefont {Ding},
  \citenamefont {Karsch},\ and\ \citenamefont {Mukherjee}}]{Ding:2015ona}%
  \BibitemOpen
  \bibfield  {author} {\bibinfo {author} {\bibfnamefont {Heng-Tong}\
  \bibnamefont {Ding}}, \bibinfo {author} {\bibfnamefont {Frithjof}\
  \bibnamefont {Karsch}}, \ and\ \bibinfo {author} {\bibfnamefont {Swagato}\
  \bibnamefont {Mukherjee}},\ }\bibfield  {title} {\enquote {\bibinfo {title}
  {{Thermodynamics of strong-interaction matter from Lattice QCD}},}\ }\href
  {\doibase 10.1142/S0218301315300076} {\bibfield  {journal} {\bibinfo
  {journal} {Int. J. Mod. Phys.}\ }\textbf {\bibinfo {volume} {E24}},\ \bibinfo
  {pages} {1530007} (\bibinfo {year} {2015})},\ \Eprint
  {http://arxiv.org/abs/1504.05274} {arXiv:1504.05274 [hep-lat]} \BibitemShut
  {NoStop}%
\bibitem [{\citenamefont {Steinbrecher}(2018)}]{Steinbrecher:2018phh}%
  \BibitemOpen
  \bibfield  {author} {\bibinfo {author} {\bibfnamefont {Patrick}\ \bibnamefont
  {Steinbrecher}},\ }\bibfield  {title} {\enquote {\bibinfo {title} {{The QCD
  crossover at zero and non-zero baryon densities from Lattice QCD}},}\
  }\href@noop {} {\  (\bibinfo {year} {2018})},\ \Eprint
  {http://arxiv.org/abs/1807.05607} {arXiv:1807.05607 [hep-lat]} \BibitemShut
  {NoStop}%
\bibitem [{\citenamefont {D'Elia}(2018)}]{DElia:2018fjp}%
  \BibitemOpen
  \bibfield  {author} {\bibinfo {author} {\bibfnamefont {Massimo}\ \bibnamefont
  {D'Elia}},\ }\bibfield  {title} {\enquote {\bibinfo {title}
  {{High-Temperature QCD: theory overview}},}\ }in\ \href@noop {} {\emph
  {\bibinfo {booktitle} {{27th International Conference on Ultrarelativistic
  Nucleus-Nucleus Collisions (Quark Matter 2018) Venice, Italy, May 14-19,
  2018}}}}\ (\bibinfo {year} {2018})\ \Eprint {http://arxiv.org/abs/1809.10660}
  {arXiv:1809.10660 [hep-lat]} \BibitemShut {NoStop}%
\bibitem [{Gee(2015)}]{Geesaman:2015fha}%
  \BibitemOpen
  \bibfield  {title} {\enquote {\bibinfo {title} {{Reaching for the horizon:
  The 2015 long range plan for nuclear science}},}\ }\href
  {http://science.energy.gov/\~/media/np/nsac/pdf/2015LRP/2015\_LRPNS\_091815.pdf}
  {\  (\bibinfo {year} {2015})}\BibitemShut {NoStop}%
\bibitem [{\citenamefont {Busza}\ \emph {et~al.}(2018)\citenamefont {Busza},
  \citenamefont {Rajagopal},\ and\ \citenamefont {van~der
  Schee}}]{Busza:2018rrf}%
  \BibitemOpen
  \bibfield  {author} {\bibinfo {author} {\bibfnamefont {Wit}\ \bibnamefont
  {Busza}}, \bibinfo {author} {\bibfnamefont {Krishna}\ \bibnamefont
  {Rajagopal}}, \ and\ \bibinfo {author} {\bibfnamefont {Wilke}\ \bibnamefont
  {van~der Schee}},\ }\bibfield  {title} {\enquote {\bibinfo {title} {{Heavy
  Ion Collisions: The Big Picture, and the Big Questions}},}\ }\href@noop {} {\
   (\bibinfo {year} {2018})},\ \Eprint {http://arxiv.org/abs/1802.04801}
  {arXiv:1802.04801 [hep-ph]} \BibitemShut {NoStop}%
\bibitem [{\citenamefont {Ablyazimov}\ \emph {et~al.}(2017)\citenamefont
  {Ablyazimov} \emph {et~al.}}]{Ablyazimov:2017guv}%
  \BibitemOpen
  \bibfield  {author} {\bibinfo {author} {\bibfnamefont {T.}~\bibnamefont
  {Ablyazimov}} \emph {et~al.} (\bibinfo {collaboration} {CBM}),\ }\bibfield
  {title} {\enquote {\bibinfo {title} {{Challenges in QCD matter physics --The
  scientific programme of the Compressed Baryonic Matter experiment at
  FAIR}},}\ }\href {\doibase 10.1140/epja/i2017-12248-y} {\bibfield  {journal}
  {\bibinfo  {journal} {Eur. Phys. J.}\ }\textbf {\bibinfo {volume} {A53}},\
  \bibinfo {pages} {60} (\bibinfo {year} {2017})},\ \Eprint
  {http://arxiv.org/abs/1607.01487} {arXiv:1607.01487 [nucl-ex]} \BibitemShut
  {NoStop}%
\bibitem [{\citenamefont {Hadjidakis}\ \emph {et~al.}(2018)\citenamefont
  {Hadjidakis} \emph {et~al.}}]{Hadjidakis:2018ifr}%
  \BibitemOpen
  \bibfield  {author} {\bibinfo {author} {\bibfnamefont {C.}~\bibnamefont
  {Hadjidakis}} \emph {et~al.},\ }\bibfield  {title} {\enquote {\bibinfo
  {title} {{A Fixed-Target Programme at the LHC: Physics Case and Projected
  Performances for Heavy-Ion, Hadron, Spin and Astroparticle Studies}},}\
  }\href@noop {} {\  (\bibinfo {year} {2018})},\ \Eprint
  {http://arxiv.org/abs/1807.00603} {arXiv:1807.00603 [hep-ex]} \BibitemShut
  {NoStop}%
\bibitem [{\citenamefont {Stephanov}(2006)}]{Stephanov:2007fk}%
  \BibitemOpen
  \bibfield  {author} {\bibinfo {author} {\bibfnamefont {M.A.}\ \bibnamefont
  {Stephanov}},\ }\bibfield  {title} {\enquote {\bibinfo {title} {{QCD phase
  .diagram: An Overview}},}\ }\href@noop {} {\bibfield  {journal} {\bibinfo
  {journal} {PoS}\ }\textbf {\bibinfo {volume} {LAT2006}},\ \bibinfo {pages}
  {024} (\bibinfo {year} {2006})},\ \Eprint
  {http://arxiv.org/abs/hep-lat/0701002} {arXiv:hep-lat/0701002 [hep-lat]}
  \BibitemShut {NoStop}%
\bibitem [{\citenamefont {Fukushima}\ and\ \citenamefont
  {Hatsuda}(2011)}]{Fukushima:2010bq}%
  \BibitemOpen
  \bibfield  {author} {\bibinfo {author} {\bibfnamefont {Kenji}\ \bibnamefont
  {Fukushima}}\ and\ \bibinfo {author} {\bibfnamefont {Tetsuo}\ \bibnamefont
  {Hatsuda}},\ }\bibfield  {title} {\enquote {\bibinfo {title} {{The phase
  diagram of dense QCD}},}\ }\href {\doibase 10.1088/0034-4885/74/1/014001}
  {\bibfield  {journal} {\bibinfo  {journal} {Rept.Prog.Phys.}\ }\textbf
  {\bibinfo {volume} {74}},\ \bibinfo {pages} {014001} (\bibinfo {year}
  {2011})},\ \Eprint {http://arxiv.org/abs/1005.4814} {arXiv:1005.4814
  [hep-ph]} \BibitemShut {NoStop}%
\bibitem [{\citenamefont {Luo}\ and\ \citenamefont {Xu}(2017)}]{Luo:2017faz}%
  \BibitemOpen
  \bibfield  {author} {\bibinfo {author} {\bibfnamefont {Xiaofeng}\
  \bibnamefont {Luo}}\ and\ \bibinfo {author} {\bibfnamefont {Nu}~\bibnamefont
  {Xu}},\ }\bibfield  {title} {\enquote {\bibinfo {title} {{Search for the QCD
  Critical Point with Fluctuations of Conserved Quantities in Relativistic
  Heavy-Ion Collisions at RHIC : An Overview}},}\ }\href {\doibase
  10.1007/s41365-017-0257-0} {\bibfield  {journal} {\bibinfo  {journal} {Nucl.
  Sci. Tech.}\ }\textbf {\bibinfo {volume} {28}},\ \bibinfo {pages} {112}
  (\bibinfo {year} {2017})},\ \Eprint {http://arxiv.org/abs/1701.02105}
  {arXiv:1701.02105 [nucl-ex]} \BibitemShut {NoStop}%
\bibitem [{\citenamefont {Videbaek}(2009)}]{Videbaek:2009zy}%
  \BibitemOpen
  \bibfield  {author} {\bibinfo {author} {\bibfnamefont {F.}~\bibnamefont
  {Videbaek}} (\bibinfo {collaboration} {BRAHMS}),\ }\bibfield  {title}
  {\enquote {\bibinfo {title} {{Overview and Recent Results from BRAHMS}},}\
  }\bibfield  {booktitle} {\emph {\bibinfo {booktitle} {{Proceedings, 21st
  International Conference on Ultra-Relativistic nucleus nucleus collisions
  (Quark matter 2009): Knoxville, USA, March 30-April 4, 2009}}},\ }\href
  {\doibase 10.1016/j.nuclphysa.2009.09.011} {\bibfield  {journal} {\bibinfo
  {journal} {Nucl. Phys.}\ }\textbf {\bibinfo {volume} {A830}},\ \bibinfo
  {pages} {43C--50C} (\bibinfo {year} {2009})},\ \Eprint
  {http://arxiv.org/abs/0907.4742} {arXiv:0907.4742 [nucl-ex]} \BibitemShut
  {NoStop}%
\bibitem [{\citenamefont {Biedron}\ and\ \citenamefont
  {Broniowski}(2007)}]{Biedron:2006vf}%
  \BibitemOpen
  \bibfield  {author} {\bibinfo {author} {\bibfnamefont {Bartlomiej}\
  \bibnamefont {Biedron}}\ and\ \bibinfo {author} {\bibfnamefont {Wojciech}\
  \bibnamefont {Broniowski}},\ }\bibfield  {title} {\enquote {\bibinfo {title}
  {{Rapidity-dependent spectra from a single-freeze-out model of relativistic
  heavy-ion collisions}},}\ }\href {\doibase 10.1103/PhysRevC.75.054905}
  {\bibfield  {journal} {\bibinfo  {journal} {Phys. Rev.}\ }\textbf {\bibinfo
  {volume} {C75}},\ \bibinfo {pages} {054905} (\bibinfo {year} {2007})},\
  \Eprint {http://arxiv.org/abs/nucl-th/0610083} {arXiv:nucl-th/0610083
  [nucl-th]} \BibitemShut {NoStop}%
\bibitem [{\citenamefont {Becattini}\ \emph {et~al.}(2007)\citenamefont
  {Becattini}, \citenamefont {Cleymans},\ and\ \citenamefont
  {Strumpfer}}]{Becattini:2007ci}%
  \BibitemOpen
  \bibfield  {author} {\bibinfo {author} {\bibfnamefont {F.}~\bibnamefont
  {Becattini}}, \bibinfo {author} {\bibfnamefont {J.}~\bibnamefont {Cleymans}},
  \ and\ \bibinfo {author} {\bibfnamefont {J.}~\bibnamefont {Strumpfer}},\
  }\bibfield  {title} {\enquote {\bibinfo {title} {{Rapidity variation of
  thermal parameters at SPS and RHIC}},}\ }\bibfield  {booktitle} {\emph
  {\bibinfo {booktitle} {{Proceedings, 4th International Workshop on Critical
  point and onset of deconfinement (CPOD07): Darmstadt, Germany, July 9-13,
  2007}}},\ }\href@noop {} {\bibfield  {journal} {\bibinfo  {journal} {PoS}\
  }\textbf {\bibinfo {volume} {CPOD07}},\ \bibinfo {pages} {012} (\bibinfo
  {year} {2007})},\ \Eprint {http://arxiv.org/abs/0709.2599} {arXiv:0709.2599
  [hep-ph]} \BibitemShut {NoStop}%
\bibitem [{\citenamefont {Begun}\ \emph {et~al.}(2018)\citenamefont {Begun},
  \citenamefont {Kikola}, \citenamefont {Vovchenko},\ and\ \citenamefont
  {Wielanek}}]{Begun:2018efg}%
  \BibitemOpen
  \bibfield  {author} {\bibinfo {author} {\bibfnamefont {Viktor}\ \bibnamefont
  {Begun}}, \bibinfo {author} {\bibfnamefont {Daniel}\ \bibnamefont {Kikola}},
  \bibinfo {author} {\bibfnamefont {Volodymyr}\ \bibnamefont {Vovchenko}}, \
  and\ \bibinfo {author} {\bibfnamefont {Daniel}\ \bibnamefont {Wielanek}},\
  }\bibfield  {title} {\enquote {\bibinfo {title} {{Estimation of the
  freeze-out parameters reachable in the AFTER@LHC project}},}\ }\href@noop {}
  {\  (\bibinfo {year} {2018})},\ \Eprint {http://arxiv.org/abs/1806.01303}
  {arXiv:1806.01303 [nucl-th]} \BibitemShut {NoStop}%
\bibitem [{\citenamefont {Li}\ and\ \citenamefont
  {Kapusta}(2017)}]{Li:2016wzh}%
  \BibitemOpen
  \bibfield  {author} {\bibinfo {author} {\bibfnamefont {Ming}\ \bibnamefont
  {Li}}\ and\ \bibinfo {author} {\bibfnamefont {Joseph~I.}\ \bibnamefont
  {Kapusta}},\ }\bibfield  {title} {\enquote {\bibinfo {title} {{High Baryon
  Densities in Heavy Ion Collisions at Energies Attainable at the BNL
  Relativistic Heavy Ion Collider and the CERN Large Hadron Collider}},}\
  }\href {\doibase 10.1103/PhysRevC.95.011901} {\bibfield  {journal} {\bibinfo
  {journal} {Phys. Rev.}\ }\textbf {\bibinfo {volume} {C95}},\ \bibinfo {pages}
  {011901} (\bibinfo {year} {2017})},\ \Eprint
  {http://arxiv.org/abs/1604.08525} {arXiv:1604.08525 [nucl-th]} \BibitemShut
  {NoStop}%
\bibitem [{\citenamefont {Shen}\ and\ \citenamefont
  {Schenke}(2018{\natexlab{a}})}]{Shen:2018pty}%
  \BibitemOpen
  \bibfield  {author} {\bibinfo {author} {\bibfnamefont {Chun}\ \bibnamefont
  {Shen}}\ and\ \bibinfo {author} {\bibfnamefont {Bj{\"o}rn}\ \bibnamefont
  {Schenke}},\ }\bibfield  {title} {\enquote {\bibinfo {title} {{Dynamical
  initialization and hydrodynamic modeling of relativistic heavy-ion
  collisions}},}\ \ }(\bibinfo {year} {2018})\ \Eprint
  {http://arxiv.org/abs/1807.05141} {arXiv:1807.05141 [nucl-th]} \BibitemShut
  {NoStop}%
\bibitem [{\citenamefont {Brewer}\ \emph {et~al.}(2018)\citenamefont {Brewer},
  \citenamefont {Mukherjee}, \citenamefont {Rajagopal},\ and\ \citenamefont
  {Yin}}]{Brewer:2018abr}%
  \BibitemOpen
  \bibfield  {author} {\bibinfo {author} {\bibfnamefont {Jasmine}\ \bibnamefont
  {Brewer}}, \bibinfo {author} {\bibfnamefont {Swagato}\ \bibnamefont
  {Mukherjee}}, \bibinfo {author} {\bibfnamefont {Krishna}\ \bibnamefont
  {Rajagopal}}, \ and\ \bibinfo {author} {\bibfnamefont {Yi}~\bibnamefont
  {Yin}},\ }\bibfield  {title} {\enquote {\bibinfo {title} {{Searching for the
  QCD critical point via the rapidity dependence of cumulants}},}\ }\href@noop
  {} {\  (\bibinfo {year} {2018})},\ \Eprint {http://arxiv.org/abs/1804.10215}
  {arXiv:1804.10215 [hep-ph]} \BibitemShut {NoStop}%
\bibitem [{\citenamefont {Stephanov}(2009)}]{Stephanov:2008qz}%
  \BibitemOpen
  \bibfield  {author} {\bibinfo {author} {\bibfnamefont {M.~A.}\ \bibnamefont
  {Stephanov}},\ }\bibfield  {title} {\enquote {\bibinfo {title} {{Non-Gaussian
  fluctuations near the QCD critical point}},}\ }\href {\doibase
  10.1103/PhysRevLett.102.032301} {\bibfield  {journal} {\bibinfo  {journal}
  {Phys. Rev. Lett.}\ }\textbf {\bibinfo {volume} {102}},\ \bibinfo {pages}
  {032301} (\bibinfo {year} {2009})},\ \Eprint {http://arxiv.org/abs/0809.3450}
  {arXiv:0809.3450 [hep-ph]} \BibitemShut {NoStop}%
\bibitem [{\citenamefont {Plumberg}\ and\ \citenamefont
  {Kapusta}(2017)}]{Plumberg:2017tvu}%
  \BibitemOpen
  \bibfield  {author} {\bibinfo {author} {\bibfnamefont {Christopher}\
  \bibnamefont {Plumberg}}\ and\ \bibinfo {author} {\bibfnamefont {Joseph~I.}\
  \bibnamefont {Kapusta}},\ }\bibfield  {title} {\enquote {\bibinfo {title}
  {{Hydrodynamic fluctuations near a critical endpoint and Hanbury-Brown?Twiss
  interferometry}},}\ }\href {\doibase 10.1103/PhysRevC.95.044910} {\bibfield
  {journal} {\bibinfo  {journal} {Phys. Rev.}\ }\textbf {\bibinfo {volume}
  {C95}},\ \bibinfo {pages} {044910} (\bibinfo {year} {2017})},\ \Eprint
  {http://arxiv.org/abs/1702.01368} {arXiv:1702.01368 [nucl-th]} \BibitemShut
  {NoStop}%
\bibitem [{\citenamefont {Bzdak}\ \emph {et~al.}(2017)\citenamefont {Bzdak},
  \citenamefont {Koch},\ and\ \citenamefont {Skokov}}]{Bzdak:2016jxo}%
  \BibitemOpen
  \bibfield  {author} {\bibinfo {author} {\bibfnamefont {Adam}\ \bibnamefont
  {Bzdak}}, \bibinfo {author} {\bibfnamefont {Volker}\ \bibnamefont {Koch}}, \
  and\ \bibinfo {author} {\bibfnamefont {Vladimir}\ \bibnamefont {Skokov}},\
  }\bibfield  {title} {\enquote {\bibinfo {title} {{Correlated stopping, proton
  clusters and higher order proton cumulants}},}\ }\href {\doibase
  10.1140/epjc/s10052-017-4847-0} {\bibfield  {journal} {\bibinfo  {journal}
  {Eur. Phys. J.}\ }\textbf {\bibinfo {volume} {C77}},\ \bibinfo {pages} {288}
  (\bibinfo {year} {2017})},\ \Eprint {http://arxiv.org/abs/1612.05128}
  {arXiv:1612.05128 [nucl-th]} \BibitemShut {NoStop}%
\bibitem [{\citenamefont {Akamatsu}\ \emph
  {et~al.}(2018{\natexlab{a}})\citenamefont {Akamatsu}, \citenamefont {Teaney},
  \citenamefont {Yan},\ and\ \citenamefont {Yin}}]{Akamatsu:2018vjr}%
  \BibitemOpen
  \bibfield  {author} {\bibinfo {author} {\bibfnamefont {Yukinao}\ \bibnamefont
  {Akamatsu}}, \bibinfo {author} {\bibfnamefont {Derek}\ \bibnamefont
  {Teaney}}, \bibinfo {author} {\bibfnamefont {Fanglida}\ \bibnamefont {Yan}},
  \ and\ \bibinfo {author} {\bibfnamefont {Yi}~\bibnamefont {Yin}},\ }\bibfield
   {title} {\enquote {\bibinfo {title} {{Transits of the QCD Critical
  Point}},}\ }\href@noop {} {\  (\bibinfo {year} {2018}{\natexlab{a}})},\
  \Eprint {http://arxiv.org/abs/1811.05081} {arXiv:1811.05081 [nucl-th]}
  \BibitemShut {NoStop}%
\bibitem [{\citenamefont {Shuryak}\ and\ \citenamefont
  {Torres-Rincon}(2018)}]{Shuryak:2018lgd}%
  \BibitemOpen
  \bibfield  {author} {\bibinfo {author} {\bibfnamefont {Edward}\ \bibnamefont
  {Shuryak}}\ and\ \bibinfo {author} {\bibfnamefont {Juan~M.}\ \bibnamefont
  {Torres-Rincon}},\ }\bibfield  {title} {\enquote {\bibinfo {title} {{Baryon
  clustering at the critical line and near the hypothetical critical point in
  heavy-ion collisions}},}\ }\href@noop {} {\  (\bibinfo {year} {2018})},\
  \Eprint {http://arxiv.org/abs/1805.04444} {arXiv:1805.04444 [hep-ph]}
  \BibitemShut {NoStop}%
\bibitem [{\citenamefont {Sun}\ \emph {et~al.}(2018)\citenamefont {Sun},
  \citenamefont {Chen}, \citenamefont {Ko}, \citenamefont {Pu},\ and\
  \citenamefont {Xu}}]{Sun:2018jhg}%
  \BibitemOpen
  \bibfield  {author} {\bibinfo {author} {\bibfnamefont {Kai-Jia}\ \bibnamefont
  {Sun}}, \bibinfo {author} {\bibfnamefont {Lie-Wen}\ \bibnamefont {Chen}},
  \bibinfo {author} {\bibfnamefont {Che~Ming}\ \bibnamefont {Ko}}, \bibinfo
  {author} {\bibfnamefont {Jie}\ \bibnamefont {Pu}}, \ and\ \bibinfo {author}
  {\bibfnamefont {Zhangbu}\ \bibnamefont {Xu}},\ }\bibfield  {title} {\enquote
  {\bibinfo {title} {{Light nuclei production as a probe of the QCD phase
  diagram}},}\ }\href {\doibase 10.1016/j.physletb.2018.04.035} {\bibfield
  {journal} {\bibinfo  {journal} {Phys. Lett.}\ }\textbf {\bibinfo {volume}
  {B781}},\ \bibinfo {pages} {499--504} (\bibinfo {year} {2018})},\ \Eprint
  {http://arxiv.org/abs/1801.09382} {arXiv:1801.09382 [nucl-th]} \BibitemShut
  {NoStop}%
\bibitem [{\citenamefont {Noronha}(2018)}]{Noronha:2018atu}%
  \BibitemOpen
  \bibfield  {author} {\bibinfo {author} {\bibfnamefont {Jorge}\ \bibnamefont
  {Noronha}},\ }\bibfield  {title} {\enquote {\bibinfo {title} {{Collective
  effects in nuclear collisions: theory overview}},}\ }in\ \href@noop {} {\emph
  {\bibinfo {booktitle} {{27th International Conference on Ultrarelativistic
  Nucleus-Nucleus Collisions (Quark Matter 2018) Venice, Italy, May 14-19,
  2018}}}}\ (\bibinfo {year} {2018})\ \Eprint {http://arxiv.org/abs/1807.06191}
  {arXiv:1807.06191 [nucl-th]} \BibitemShut {NoStop}%
\bibitem [{\citenamefont {Shen}\ and\ \citenamefont
  {Schenke}(2018{\natexlab{b}})}]{Shen:2017bsr}%
  \BibitemOpen
  \bibfield  {author} {\bibinfo {author} {\bibfnamefont {Chun}\ \bibnamefont
  {Shen}}\ and\ \bibinfo {author} {\bibfnamefont {Bjorn}\ \bibnamefont
  {Schenke}},\ }\bibfield  {title} {\enquote {\bibinfo {title} {{Dynamical
  initial state model for relativistic heavy-ion collisions}},}\ }\href
  {\doibase 10.1103/PhysRevC.97.024907} {\bibfield  {journal} {\bibinfo
  {journal} {Phys. Rev.}\ }\textbf {\bibinfo {volume} {C97}},\ \bibinfo {pages}
  {024907} (\bibinfo {year} {2018}{\natexlab{b}})},\ \Eprint
  {http://arxiv.org/abs/1710.00881} {arXiv:1710.00881 [nucl-th]} \BibitemShut
  {NoStop}%
\bibitem [{\citenamefont {Okai}\ \emph {et~al.}(2017)\citenamefont {Okai},
  \citenamefont {Kawaguchi}, \citenamefont {Tachibana},\ and\ \citenamefont
  {Hirano}}]{Okai:2017ofp}%
  \BibitemOpen
  \bibfield  {author} {\bibinfo {author} {\bibfnamefont {Michito}\ \bibnamefont
  {Okai}}, \bibinfo {author} {\bibfnamefont {Koji}\ \bibnamefont {Kawaguchi}},
  \bibinfo {author} {\bibfnamefont {Yasuki}\ \bibnamefont {Tachibana}}, \ and\
  \bibinfo {author} {\bibfnamefont {Tetsufumi}\ \bibnamefont {Hirano}},\
  }\bibfield  {title} {\enquote {\bibinfo {title} {{New approach to
  initializing hydrodynamic fields and mini-jet propagation in quark-gluon
  fluids}},}\ }\href {\doibase 10.1103/PhysRevC.95.054914} {\bibfield
  {journal} {\bibinfo  {journal} {Phys. Rev.}\ }\textbf {\bibinfo {volume}
  {C95}},\ \bibinfo {pages} {054914} (\bibinfo {year} {2017})},\ \Eprint
  {http://arxiv.org/abs/1702.07541} {arXiv:1702.07541 [nucl-th]} \BibitemShut
  {NoStop}%
\bibitem [{\citenamefont {Akamatsu}\ \emph
  {et~al.}(2018{\natexlab{b}})\citenamefont {Akamatsu}, \citenamefont
  {Asakawa}, \citenamefont {Hirano}, \citenamefont {Kitazawa}, \citenamefont
  {Morita}, \citenamefont {Murase}, \citenamefont {Nara}, \citenamefont
  {Nonaka},\ and\ \citenamefont {Ohnishi}}]{Akamatsu:2018olk}%
  \BibitemOpen
  \bibfield  {author} {\bibinfo {author} {\bibfnamefont {Yukinao}\ \bibnamefont
  {Akamatsu}}, \bibinfo {author} {\bibfnamefont {Masayuki}\ \bibnamefont
  {Asakawa}}, \bibinfo {author} {\bibfnamefont {Tetsufumi}\ \bibnamefont
  {Hirano}}, \bibinfo {author} {\bibfnamefont {Masakiyo}\ \bibnamefont
  {Kitazawa}}, \bibinfo {author} {\bibfnamefont {Kenji}\ \bibnamefont
  {Morita}}, \bibinfo {author} {\bibfnamefont {Koichi}\ \bibnamefont {Murase}},
  \bibinfo {author} {\bibfnamefont {Yasushi}\ \bibnamefont {Nara}}, \bibinfo
  {author} {\bibfnamefont {Chiho}\ \bibnamefont {Nonaka}}, \ and\ \bibinfo
  {author} {\bibfnamefont {Akira}\ \bibnamefont {Ohnishi}},\ }\bibfield
  {title} {\enquote {\bibinfo {title} {{A dynamically integrated transport
  approach for heavy-ion collisions at high baryon density}},}\ }\href@noop {}
  {\  (\bibinfo {year} {2018}{\natexlab{b}})},\ \Eprint
  {http://arxiv.org/abs/1805.09024} {arXiv:1805.09024 [nucl-th]} \BibitemShut
  {NoStop}%
\bibitem [{\citenamefont {Denicol}\ \emph {et~al.}(2018)\citenamefont
  {Denicol}, \citenamefont {Gale}, \citenamefont {Jeon}, \citenamefont
  {Monnai}, \citenamefont {Schenke},\ and\ \citenamefont
  {Shen}}]{Denicol:2018wdp}%
  \BibitemOpen
  \bibfield  {author} {\bibinfo {author} {\bibfnamefont {Gabriel~S.}\
  \bibnamefont {Denicol}}, \bibinfo {author} {\bibfnamefont {Charles}\
  \bibnamefont {Gale}}, \bibinfo {author} {\bibfnamefont {Sangyong}\
  \bibnamefont {Jeon}}, \bibinfo {author} {\bibfnamefont {Akihiko}\
  \bibnamefont {Monnai}}, \bibinfo {author} {\bibfnamefont {Bj{\"o}rn}\
  \bibnamefont {Schenke}}, \ and\ \bibinfo {author} {\bibfnamefont {Chun}\
  \bibnamefont {Shen}},\ }\bibfield  {title} {\enquote {\bibinfo {title} {{Net
  baryon diffusion in fluid dynamic simulations of relativistic heavy-ion
  collisions}},}\ }\href@noop {} {\  (\bibinfo {year} {2018})},\ \Eprint
  {http://arxiv.org/abs/1804.10557} {arXiv:1804.10557 [nucl-th]} \BibitemShut
  {NoStop}%
\bibitem [{\citenamefont {Du}\ \emph {et~al.}(2018)\citenamefont {Du},
  \citenamefont {Heinz},\ and\ \citenamefont {Vujanovic}}]{Du:2018mpf}%
  \BibitemOpen
  \bibfield  {author} {\bibinfo {author} {\bibfnamefont {Lipei}\ \bibnamefont
  {Du}}, \bibinfo {author} {\bibfnamefont {Ulrich}\ \bibnamefont {Heinz}}, \
  and\ \bibinfo {author} {\bibfnamefont {Gojko}\ \bibnamefont {Vujanovic}},\
  }\bibfield  {title} {\enquote {\bibinfo {title} {{Hybrid model with dynamical
  sources for heavy-ion collisions at BES energies}},}\ }\href@noop {} {\
  (\bibinfo {year} {2018})},\ \Eprint {http://arxiv.org/abs/1807.04721}
  {arXiv:1807.04721 [nucl-th]} \BibitemShut {NoStop}%
\bibitem [{\citenamefont {Petersen}(2017)}]{Petersen:2017jdb}%
  \BibitemOpen
  \bibfield  {author} {\bibinfo {author} {\bibfnamefont {H.}~\bibnamefont
  {Petersen}},\ }\bibfield  {title} {\enquote {\bibinfo {title} {{Beam energy
  scan theory: Status and open questions}},}\ }\bibfield  {booktitle} {\emph
  {\bibinfo {booktitle} {{Proceedings, 26th International Conference on
  Ultra-relativistic Nucleus-Nucleus Collisions (Quark Matter 2017): Chicago,
  Illinois, USA, February 5-11, 2017}}},\ }\href {\doibase
  10.1016/j.nuclphysa.2017.04.018} {\bibfield  {journal} {\bibinfo  {journal}
  {Nucl. Phys.}\ }\textbf {\bibinfo {volume} {A967}},\ \bibinfo {pages}
  {145--152} (\bibinfo {year} {2017})},\ \Eprint
  {http://arxiv.org/abs/1704.02904} {arXiv:1704.02904 [nucl-th]} \BibitemShut
  {NoStop}%
\bibitem [{\citenamefont {Parotto}\ \emph {et~al.}(2018)\citenamefont
  {Parotto}, \citenamefont {Bluhm}, \citenamefont {Mroczek}, \citenamefont
  {Nahrgang}, \citenamefont {Noronha-Hostler}, \citenamefont {Rajagopal},
  \citenamefont {Ratti}, \citenamefont {Sch{\"a}fer},\ and\ \citenamefont
  {Stephanov}}]{Parotto:2018pwx}%
  \BibitemOpen
  \bibfield  {author} {\bibinfo {author} {\bibfnamefont {Paolo}\ \bibnamefont
  {Parotto}}, \bibinfo {author} {\bibfnamefont {Marcus}\ \bibnamefont {Bluhm}},
  \bibinfo {author} {\bibfnamefont {Debora}\ \bibnamefont {Mroczek}}, \bibinfo
  {author} {\bibfnamefont {Marlene}\ \bibnamefont {Nahrgang}}, \bibinfo
  {author} {\bibfnamefont {Jacquelyn}\ \bibnamefont {Noronha-Hostler}},
  \bibinfo {author} {\bibfnamefont {Krishna}\ \bibnamefont {Rajagopal}},
  \bibinfo {author} {\bibfnamefont {Claudia}\ \bibnamefont {Ratti}}, \bibinfo
  {author} {\bibfnamefont {Thomas}\ \bibnamefont {Sch{\"a}fer}}, \ and\
  \bibinfo {author} {\bibfnamefont {Mikhail}\ \bibnamefont {Stephanov}},\
  }\bibfield  {title} {\enquote {\bibinfo {title} {{Lattice-QCD-based equation
  of state with a critical point}},}\ }\href@noop {} {\  (\bibinfo {year}
  {2018})},\ \Eprint {http://arxiv.org/abs/1805.05249} {arXiv:1805.05249
  [hep-ph]} \BibitemShut {NoStop}%
\bibitem [{\citenamefont {Nonaka}\ and\ \citenamefont
  {Asakawa}(2005)}]{Nonaka:2004pg}%
  \BibitemOpen
  \bibfield  {author} {\bibinfo {author} {\bibfnamefont {Chiho}\ \bibnamefont
  {Nonaka}}\ and\ \bibinfo {author} {\bibfnamefont {Masayuki}\ \bibnamefont
  {Asakawa}},\ }\bibfield  {title} {\enquote {\bibinfo {title} {{Hydrodynamical
  evolution near the QCD critical end point}},}\ }\href {\doibase
  10.1103/PhysRevC.71.044904} {\bibfield  {journal} {\bibinfo  {journal} {Phys.
  Rev.}\ }\textbf {\bibinfo {volume} {C71}},\ \bibinfo {pages} {044904}
  (\bibinfo {year} {2005})},\ \Eprint {http://arxiv.org/abs/nucl-th/0410078}
  {arXiv:nucl-th/0410078 [nucl-th]} \BibitemShut {NoStop}%
\bibitem [{\citenamefont {Kapusta}\ and\ \citenamefont
  {Torres-Rincon}(2012)}]{Kapusta:2012zb}%
  \BibitemOpen
  \bibfield  {author} {\bibinfo {author} {\bibfnamefont {Joseph~I.}\
  \bibnamefont {Kapusta}}\ and\ \bibinfo {author} {\bibfnamefont {Juan~M.}\
  \bibnamefont {Torres-Rincon}},\ }\bibfield  {title} {\enquote {\bibinfo
  {title} {{Thermal Conductivity and Chiral Critical Point in Heavy Ion
  Collisions}},}\ }\href {\doibase 10.1103/PhysRevC.86.054911} {\bibfield
  {journal} {\bibinfo  {journal} {Phys. Rev.}\ }\textbf {\bibinfo {volume}
  {C86}},\ \bibinfo {pages} {054911} (\bibinfo {year} {2012})},\ \Eprint
  {http://arxiv.org/abs/1209.0675} {arXiv:1209.0675 [nucl-th]} \BibitemShut
  {NoStop}%
\bibitem [{\citenamefont {Monnai}\ \emph {et~al.}(2017)\citenamefont {Monnai},
  \citenamefont {Mukherjee},\ and\ \citenamefont {Yin}}]{Monnai:2016kud}%
  \BibitemOpen
  \bibfield  {author} {\bibinfo {author} {\bibfnamefont {Akihiko}\ \bibnamefont
  {Monnai}}, \bibinfo {author} {\bibfnamefont {Swagato}\ \bibnamefont
  {Mukherjee}}, \ and\ \bibinfo {author} {\bibfnamefont {Yi}~\bibnamefont
  {Yin}},\ }\bibfield  {title} {\enquote {\bibinfo {title} {{Phenomenological
  Consequences of Enhanced Bulk Viscosity Near the QCD Critical Point}},}\
  }\href {\doibase 10.1103/PhysRevC.95.034902} {\bibfield  {journal} {\bibinfo
  {journal} {Phys. Rev.}\ }\textbf {\bibinfo {volume} {C95}},\ \bibinfo {pages}
  {034902} (\bibinfo {year} {2017})},\ \Eprint
  {http://arxiv.org/abs/1606.00771} {arXiv:1606.00771 [nucl-th]} \BibitemShut
  {NoStop}%
\bibitem [{\citenamefont {Onuki}(2002)}]{onuki2002phase}%
  \BibitemOpen
  \bibfield  {author} {\bibinfo {author} {\bibfnamefont {A.}~\bibnamefont
  {Onuki}},\ }\href {https://books.google.com/books?id=DA6AAVfv11sC} {\emph
  {\bibinfo {title} {Phase Transition Dynamics}}}\ (\bibinfo  {publisher}
  {Cambridge University Press},\ \bibinfo {year} {2002})\BibitemShut {NoStop}%
\bibitem [{\citenamefont {Luo}(2016)}]{Luo:2015doi}%
  \BibitemOpen
  \bibfield  {author} {\bibinfo {author} {\bibfnamefont {Xiaofeng}\
  \bibnamefont {Luo}},\ }\bibfield  {title} {\enquote {\bibinfo {title}
  {{Exploring the QCD Phase Structure with Beam Energy Scan in Heavy-ion
  Collisions}},}\ }\bibfield  {booktitle} {\emph {\bibinfo {booktitle}
  {{Proceedings, 25th International Conference on Ultra-Relativistic
  Nucleus-Nucleus Collisions (Quark Matter 2015): Kobe, Japan, September
  27-October 3, 2015}}},\ }\href {\doibase 10.1016/j.nuclphysa.2016.03.025}
  {\bibfield  {journal} {\bibinfo  {journal} {Nucl. Phys.}\ }\textbf {\bibinfo
  {volume} {A956}},\ \bibinfo {pages} {75--82} (\bibinfo {year} {2016})},\
  \Eprint {http://arxiv.org/abs/1512.09215} {arXiv:1512.09215 [nucl-ex]}
  \BibitemShut {NoStop}%
\bibitem [{\citenamefont {Jiang}\ \emph {et~al.}(2016)\citenamefont {Jiang},
  \citenamefont {Li},\ and\ \citenamefont {Song}}]{Jiang:2015hri}%
  \BibitemOpen
  \bibfield  {author} {\bibinfo {author} {\bibfnamefont {Lijia}\ \bibnamefont
  {Jiang}}, \bibinfo {author} {\bibfnamefont {Pengfei}\ \bibnamefont {Li}}, \
  and\ \bibinfo {author} {\bibfnamefont {Huichao}\ \bibnamefont {Song}},\
  }\bibfield  {title} {\enquote {\bibinfo {title} {{Correlated fluctuations
  near the QCD critical point}},}\ }\href {\doibase 10.1103/PhysRevC.94.024918}
  {\bibfield  {journal} {\bibinfo  {journal} {Phys. Rev.}\ }\textbf {\bibinfo
  {volume} {C94}},\ \bibinfo {pages} {024918} (\bibinfo {year} {2016})},\
  \Eprint {http://arxiv.org/abs/1512.06164} {arXiv:1512.06164 [nucl-th]}
  \BibitemShut {NoStop}%
\bibitem [{\citenamefont {Mukherjee}\ \emph {et~al.}(2015)\citenamefont
  {Mukherjee}, \citenamefont {Venugopalan},\ and\ \citenamefont
  {Yin}}]{Mukherjee:2015swa}%
  \BibitemOpen
  \bibfield  {author} {\bibinfo {author} {\bibfnamefont {Swagato}\ \bibnamefont
  {Mukherjee}}, \bibinfo {author} {\bibfnamefont {Raju}\ \bibnamefont
  {Venugopalan}}, \ and\ \bibinfo {author} {\bibfnamefont {Yi}~\bibnamefont
  {Yin}},\ }\bibfield  {title} {\enquote {\bibinfo {title} {{Real time
  evolution of non-Gaussian cumulants in the QCD critical regime}},}\ }\href
  {\doibase 10.1103/PhysRevC.92.034912} {\bibfield  {journal} {\bibinfo
  {journal} {Phys. Rev.}\ }\textbf {\bibinfo {volume} {C92}},\ \bibinfo {pages}
  {034912} (\bibinfo {year} {2015})},\ \Eprint
  {http://arxiv.org/abs/1506.00645} {arXiv:1506.00645 [hep-ph]} \BibitemShut
  {NoStop}%
\bibitem [{\citenamefont {Nahrgang}\ \emph {et~al.}(2018)\citenamefont
  {Nahrgang}, \citenamefont {Bluhm}, \citenamefont {Sch{\"a}fer},\ and\
  \citenamefont {Bass}}]{Nahrgang:2018afz}%
  \BibitemOpen
  \bibfield  {author} {\bibinfo {author} {\bibfnamefont {Marlene}\ \bibnamefont
  {Nahrgang}}, \bibinfo {author} {\bibfnamefont {Marcus}\ \bibnamefont
  {Bluhm}}, \bibinfo {author} {\bibfnamefont {Thomas}\ \bibnamefont
  {Sch{\"a}fer}}, \ and\ \bibinfo {author} {\bibfnamefont {Steffen~A.}\
  \bibnamefont {Bass}},\ }\bibfield  {title} {\enquote {\bibinfo {title}
  {{Diffusive dynamics of critical fluctuations near the QCD critical
  point}},}\ }\href@noop {} {\  (\bibinfo {year} {2018})},\ \Eprint
  {http://arxiv.org/abs/1804.05728} {arXiv:1804.05728 [nucl-th]} \BibitemShut
  {NoStop}%
\bibitem [{\citenamefont {Zurek}(1996)}]{Zurek:1996sj}%
  \BibitemOpen
  \bibfield  {author} {\bibinfo {author} {\bibfnamefont {W.~H.}\ \bibnamefont
  {Zurek}},\ }\bibfield  {title} {\enquote {\bibinfo {title} {{Cosmological
  experiments in condensed matter systems}},}\ }\href {\doibase
  10.1016/S0370-1573(96)00009-9} {\bibfield  {journal} {\bibinfo  {journal}
  {Phys. Rept.}\ }\textbf {\bibinfo {volume} {276}},\ \bibinfo {pages}
  {177--221} (\bibinfo {year} {1996})},\ \Eprint
  {http://arxiv.org/abs/cond-mat/9607135} {arXiv:cond-mat/9607135 [cond-mat]}
  \BibitemShut {NoStop}%
\bibitem [{\citenamefont {Mukherjee}\ \emph {et~al.}(2016)\citenamefont
  {Mukherjee}, \citenamefont {Venugopalan},\ and\ \citenamefont
  {Yin}}]{Mukherjee:2016kyu}%
  \BibitemOpen
  \bibfield  {author} {\bibinfo {author} {\bibfnamefont {Swagato}\ \bibnamefont
  {Mukherjee}}, \bibinfo {author} {\bibfnamefont {Raju}\ \bibnamefont
  {Venugopalan}}, \ and\ \bibinfo {author} {\bibfnamefont {Yi}~\bibnamefont
  {Yin}},\ }\bibfield  {title} {\enquote {\bibinfo {title} {{Universal
  off-equilibrium scaling of critical cumulants in the QCD phase diagram}},}\
  }\href {\doibase 10.1103/PhysRevLett.117.222301} {\bibfield  {journal}
  {\bibinfo  {journal} {Phys. Rev. Lett.}\ }\textbf {\bibinfo {volume} {117}},\
  \bibinfo {pages} {222301} (\bibinfo {year} {2016})},\ \Eprint
  {http://arxiv.org/abs/1605.09341} {arXiv:1605.09341 [hep-ph]} \BibitemShut
  {NoStop}%
\bibitem [{\citenamefont {Griffin}\ \emph {et~al.}(2012)\citenamefont
  {Griffin}, \citenamefont {Lilienblum}, \citenamefont {Delaney}, \citenamefont
  {Kumagai}, \citenamefont {Fiebig},\ and\ \citenamefont
  {Spaldin}}]{PhysRevX.2.041022}%
  \BibitemOpen
  \bibfield  {author} {\bibinfo {author} {\bibfnamefont {S.~M.}\ \bibnamefont
  {Griffin}}, \bibinfo {author} {\bibfnamefont {M.}~\bibnamefont {Lilienblum}},
  \bibinfo {author} {\bibfnamefont {K.~T.}\ \bibnamefont {Delaney}}, \bibinfo
  {author} {\bibfnamefont {Y.}~\bibnamefont {Kumagai}}, \bibinfo {author}
  {\bibfnamefont {M.}~\bibnamefont {Fiebig}}, \ and\ \bibinfo {author}
  {\bibfnamefont {N.~A.}\ \bibnamefont {Spaldin}},\ }\bibfield  {title}
  {\enquote {\bibinfo {title} {Scaling behavior and beyond equilibrium in the
  hexagonal manganites},}\ }\href {\doibase 10.1103/PhysRevX.2.041022}
  {\bibfield  {journal} {\bibinfo  {journal} {Phys. Rev. X}\ }\textbf {\bibinfo
  {volume} {2}},\ \bibinfo {pages} {041022} (\bibinfo {year}
  {2012})}\BibitemShut {NoStop}%
\bibitem [{\citenamefont {Clark}\ \emph {et~al.}(2016)\citenamefont {Clark},
  \citenamefont {Feng},\ and\ \citenamefont {Chin}}]{Clark606}%
  \BibitemOpen
  \bibfield  {author} {\bibinfo {author} {\bibfnamefont {Logan~W.}\
  \bibnamefont {Clark}}, \bibinfo {author} {\bibfnamefont {Lei}\ \bibnamefont
  {Feng}}, \ and\ \bibinfo {author} {\bibfnamefont {Cheng}\ \bibnamefont
  {Chin}},\ }\bibfield  {title} {\enquote {\bibinfo {title} {Universal
  space-time scaling symmetry in the dynamics of bosons across a quantum phase
  transition},}\ }\href {\doibase 10.1126/science.aaf9657} {\bibfield
  {journal} {\bibinfo  {journal} {Science}\ }\textbf {\bibinfo {volume}
  {354}},\ \bibinfo {pages} {606--610} (\bibinfo {year} {2016})},\ \Eprint
  {http://arxiv.org/abs/http://science.sciencemag.org/content/354/6312/606.full.pdf}
  {http://science.sciencemag.org/content/354/6312/606.full.pdf} \BibitemShut
  {NoStop}%
\bibitem [{\citenamefont {Berdnikov}\ and\ \citenamefont
  {Rajagopal}(2000)}]{Berdnikov:1999ph}%
  \BibitemOpen
  \bibfield  {author} {\bibinfo {author} {\bibfnamefont {Boris}\ \bibnamefont
  {Berdnikov}}\ and\ \bibinfo {author} {\bibfnamefont {Krishna}\ \bibnamefont
  {Rajagopal}},\ }\bibfield  {title} {\enquote {\bibinfo {title} {{Slowing
  out-of-equilibrium near the QCD critical point}},}\ }\href {\doibase
  10.1103/PhysRevD.61.105017} {\bibfield  {journal} {\bibinfo  {journal} {Phys.
  Rev.}\ }\textbf {\bibinfo {volume} {D61}},\ \bibinfo {pages} {105017}
  (\bibinfo {year} {2000})},\ \Eprint {http://arxiv.org/abs/hep-ph/9912274}
  {arXiv:hep-ph/9912274 [hep-ph]} \BibitemShut {NoStop}%
\bibitem [{\citenamefont {Herold}\ \emph {et~al.}(2013)\citenamefont {Herold},
  \citenamefont {Nahrgang}, \citenamefont {Mishustin},\ and\ \citenamefont
  {Bleicher}}]{Herold:2013bi}%
  \BibitemOpen
  \bibfield  {author} {\bibinfo {author} {\bibfnamefont {Christoph}\
  \bibnamefont {Herold}}, \bibinfo {author} {\bibfnamefont {Marlene}\
  \bibnamefont {Nahrgang}}, \bibinfo {author} {\bibfnamefont {Igor}\
  \bibnamefont {Mishustin}}, \ and\ \bibinfo {author} {\bibfnamefont {Marcus}\
  \bibnamefont {Bleicher}},\ }\bibfield  {title} {\enquote {\bibinfo {title}
  {{Chiral fluid dynamics with explicit propagation of the Polyakov loop}},}\
  }\href {\doibase 10.1103/PhysRevC.87.014907} {\bibfield  {journal} {\bibinfo
  {journal} {Phys. Rev.}\ }\textbf {\bibinfo {volume} {C87}},\ \bibinfo {pages}
  {014907} (\bibinfo {year} {2013})},\ \Eprint {http://arxiv.org/abs/1301.1214}
  {arXiv:1301.1214 [nucl-th]} \BibitemShut {NoStop}%
\bibitem [{\citenamefont {Herold}\ \emph {et~al.}(2014)\citenamefont {Herold},
  \citenamefont {Nahrgang}, \citenamefont {Yan},\ and\ \citenamefont
  {Kobdaj}}]{Herold:2014zoa}%
  \BibitemOpen
  \bibfield  {author} {\bibinfo {author} {\bibfnamefont {Christoph}\
  \bibnamefont {Herold}}, \bibinfo {author} {\bibfnamefont {Marlene}\
  \bibnamefont {Nahrgang}}, \bibinfo {author} {\bibfnamefont {Yupeng}\
  \bibnamefont {Yan}}, \ and\ \bibinfo {author} {\bibfnamefont {Chinorat}\
  \bibnamefont {Kobdaj}},\ }\bibfield  {title} {\enquote {\bibinfo {title}
  {{Net-baryon number variance and kurtosis within nonequilibrium chiral fluid
  dynamics}},}\ }\href {\doibase 10.1088/0954-3899/41/11/115106} {\bibfield
  {journal} {\bibinfo  {journal} {J. Phys.}\ }\textbf {\bibinfo {volume}
  {G41}},\ \bibinfo {pages} {115106} (\bibinfo {year} {2014})},\ \Eprint
  {http://arxiv.org/abs/1407.8277} {arXiv:1407.8277 [hep-ph]} \BibitemShut
  {NoStop}%
\bibitem [{\citenamefont {Herold}\ \emph {et~al.}(2016)\citenamefont {Herold},
  \citenamefont {Nahrgang}, \citenamefont {Yan},\ and\ \citenamefont
  {Kobdaj}}]{Herold:2016uvv}%
  \BibitemOpen
  \bibfield  {author} {\bibinfo {author} {\bibfnamefont {Christoph}\
  \bibnamefont {Herold}}, \bibinfo {author} {\bibfnamefont {Marlene}\
  \bibnamefont {Nahrgang}}, \bibinfo {author} {\bibfnamefont {Yupeng}\
  \bibnamefont {Yan}}, \ and\ \bibinfo {author} {\bibfnamefont {Chinorat}\
  \bibnamefont {Kobdaj}},\ }\bibfield  {title} {\enquote {\bibinfo {title}
  {{Dynamical net-proton fluctuations near a QCD critical point}},}\ }\href
  {\doibase 10.1103/PhysRevC.93.021902} {\bibfield  {journal} {\bibinfo
  {journal} {Phys. Rev.}\ }\textbf {\bibinfo {volume} {C93}},\ \bibinfo {pages}
  {021902} (\bibinfo {year} {2016})},\ \Eprint
  {http://arxiv.org/abs/1601.04839} {arXiv:1601.04839 [hep-ph]} \BibitemShut
  {NoStop}%
\bibitem [{\citenamefont {Nahrgang}(2016)}]{Nahrgang:2016ayr}%
  \BibitemOpen
  \bibfield  {author} {\bibinfo {author} {\bibfnamefont {Marlene}\ \bibnamefont
  {Nahrgang}},\ }\bibfield  {title} {\enquote {\bibinfo {title} {{The QCD
  Critical Point and Related Observables}},}\ }\bibfield  {booktitle} {\emph
  {\bibinfo {booktitle} {{Proceedings, 25th International Conference on
  Ultra-Relativistic Nucleus-Nucleus Collisions (Quark Matter 2015): Kobe,
  Japan, September 27-October 3, 2015}}},\ }\href {\doibase
  10.1016/j.nuclphysa.2016.02.074} {\bibfield  {journal} {\bibinfo  {journal}
  {Nucl. Phys.}\ }\textbf {\bibinfo {volume} {A956}},\ \bibinfo {pages}
  {83--90} (\bibinfo {year} {2016})},\ \Eprint
  {http://arxiv.org/abs/1601.07437} {arXiv:1601.07437 [nucl-th]} \BibitemShut
  {NoStop}%
\bibitem [{\citenamefont {Son}\ and\ \citenamefont
  {Stephanov}(2004)}]{Son:2004iv}%
  \BibitemOpen
  \bibfield  {author} {\bibinfo {author} {\bibfnamefont {D.~T.}\ \bibnamefont
  {Son}}\ and\ \bibinfo {author} {\bibfnamefont {M.~A.}\ \bibnamefont
  {Stephanov}},\ }\bibfield  {title} {\enquote {\bibinfo {title} {{Dynamic
  universality class of the QCD critical point}},}\ }\href {\doibase
  10.1103/PhysRevD.70.056001} {\bibfield  {journal} {\bibinfo  {journal} {Phys.
  Rev.}\ }\textbf {\bibinfo {volume} {D70}},\ \bibinfo {pages} {056001}
  (\bibinfo {year} {2004})},\ \Eprint {http://arxiv.org/abs/hep-ph/0401052}
  {arXiv:hep-ph/0401052 [hep-ph]} \BibitemShut {NoStop}%
\bibitem [{\citenamefont {Fujii}\ and\ \citenamefont
  {Ohtani}(2004)}]{Fujii:2004jt}%
  \BibitemOpen
  \bibfield  {author} {\bibinfo {author} {\bibfnamefont {H.}~\bibnamefont
  {Fujii}}\ and\ \bibinfo {author} {\bibfnamefont {M.}~\bibnamefont {Ohtani}},\
  }\bibfield  {title} {\enquote {\bibinfo {title} {{Sigma and hydrodynamic
  modes along the critical line}},}\ }\href {\doibase
  10.1103/PhysRevD.70.014016} {\bibfield  {journal} {\bibinfo  {journal} {Phys.
  Rev.}\ }\textbf {\bibinfo {volume} {D70}},\ \bibinfo {pages} {014016}
  (\bibinfo {year} {2004})},\ \Eprint {http://arxiv.org/abs/hep-ph/0402263}
  {arXiv:hep-ph/0402263 [hep-ph]} \BibitemShut {NoStop}%
\bibitem [{\citenamefont {Landau}\ and\ \citenamefont
  {Lifshitz}(2013)}]{landau2013fluid}%
  \BibitemOpen
  \bibfield  {author} {\bibinfo {author} {\bibfnamefont {L.D.}\ \bibnamefont
  {Landau}}\ and\ \bibinfo {author} {\bibfnamefont {E.M.}\ \bibnamefont
  {Lifshitz}},\ }\href@noop {} {\emph {\bibinfo {title} {Fluid Mechanics:
  Landau and Lifshitz: Course of Theoretical Physics}}},\ Vol.~\bibinfo
  {volume} {6}\ (\bibinfo  {publisher} {Elsevier Science},\ \bibinfo {year}
  {2013})\BibitemShut {NoStop}%
\bibitem [{\citenamefont {Kapusta}\ \emph {et~al.}(2012)\citenamefont
  {Kapusta}, \citenamefont {Muller},\ and\ \citenamefont
  {Stephanov}}]{Kapusta:2011gt}%
  \BibitemOpen
  \bibfield  {author} {\bibinfo {author} {\bibfnamefont {J.~I.}\ \bibnamefont
  {Kapusta}}, \bibinfo {author} {\bibfnamefont {B.}~\bibnamefont {Muller}}, \
  and\ \bibinfo {author} {\bibfnamefont {M.}~\bibnamefont {Stephanov}},\
  }\bibfield  {title} {\enquote {\bibinfo {title} {{Relativistic Theory of
  Hydrodynamic Fluctuations with Applications to Heavy Ion Collisions}},}\
  }\href {\doibase 10.1103/PhysRevC.85.054906} {\bibfield  {journal} {\bibinfo
  {journal} {Phys. Rev.}\ }\textbf {\bibinfo {volume} {C85}},\ \bibinfo {pages}
  {054906} (\bibinfo {year} {2012})},\ \Eprint {http://arxiv.org/abs/1112.6405}
  {arXiv:1112.6405 [nucl-th]} \BibitemShut {NoStop}%
\bibitem [{\citenamefont {Murase}\ and\ \citenamefont
  {Hirano}(2013)}]{Murase:2013tma}%
  \BibitemOpen
  \bibfield  {author} {\bibinfo {author} {\bibfnamefont {Koichi}\ \bibnamefont
  {Murase}}\ and\ \bibinfo {author} {\bibfnamefont {Tetsufumi}\ \bibnamefont
  {Hirano}},\ }\bibfield  {title} {\enquote {\bibinfo {title} {{Relativistic
  fluctuating hydrodynamics with memory functions and colored noises}},}\
  }\href@noop {} {\  (\bibinfo {year} {2013})},\ \Eprint
  {http://arxiv.org/abs/1304.3243} {arXiv:1304.3243 [nucl-th]} \BibitemShut
  {NoStop}%
\bibitem [{\citenamefont {Singh}\ \emph {et~al.}(2018)\citenamefont {Singh},
  \citenamefont {Shen}, \citenamefont {McDonald}, \citenamefont {Jeon},\ and\
  \citenamefont {Gale}}]{Singh:2018dpk}%
  \BibitemOpen
  \bibfield  {author} {\bibinfo {author} {\bibfnamefont {Mayank}\ \bibnamefont
  {Singh}}, \bibinfo {author} {\bibfnamefont {Chun}\ \bibnamefont {Shen}},
  \bibinfo {author} {\bibfnamefont {Scott}\ \bibnamefont {McDonald}}, \bibinfo
  {author} {\bibfnamefont {Sangyong}\ \bibnamefont {Jeon}}, \ and\ \bibinfo
  {author} {\bibfnamefont {Charles}\ \bibnamefont {Gale}},\ }\bibfield  {title}
  {\enquote {\bibinfo {title} {{Hydrodynamic Fluctuations in Relativistic
  Heavy-Ion Collisions}},}\ }in\ \href@noop {} {\emph {\bibinfo {booktitle}
  {{27th International Conference on Ultrarelativistic Nucleus-Nucleus
  Collisions (Quark Matter 2018) Venice, Italy, May 14-19, 2018}}}}\ (\bibinfo
  {year} {2018})\ \Eprint {http://arxiv.org/abs/1807.05451} {arXiv:1807.05451
  [nucl-th]} \BibitemShut {NoStop}%
\bibitem [{\citenamefont {Sakaida}\ \emph {et~al.}(2017)\citenamefont
  {Sakaida}, \citenamefont {Asakawa}, \citenamefont {Fujii},\ and\
  \citenamefont {Kitazawa}}]{Sakaida:2017rtj}%
  \BibitemOpen
  \bibfield  {author} {\bibinfo {author} {\bibfnamefont {Miki}\ \bibnamefont
  {Sakaida}}, \bibinfo {author} {\bibfnamefont {Masayuki}\ \bibnamefont
  {Asakawa}}, \bibinfo {author} {\bibfnamefont {Hirotsugu}\ \bibnamefont
  {Fujii}}, \ and\ \bibinfo {author} {\bibfnamefont {Masakiyo}\ \bibnamefont
  {Kitazawa}},\ }\bibfield  {title} {\enquote {\bibinfo {title} {{Dynamical
  evolution of critical fluctuations and its observation in heavy ion
  collisions}},}\ }\href {\doibase 10.1103/PhysRevC.95.064905} {\bibfield
  {journal} {\bibinfo  {journal} {Phys. Rev.}\ }\textbf {\bibinfo {volume}
  {C95}},\ \bibinfo {pages} {064905} (\bibinfo {year} {2017})},\ \Eprint
  {http://arxiv.org/abs/1703.08008} {arXiv:1703.08008 [nucl-th]} \BibitemShut
  {NoStop}%
\bibitem [{\citenamefont {Lebedev}\ \emph {et~al.}(1983)\citenamefont
  {Lebedev}, \citenamefont {Sukhorukov},\ and\ \citenamefont
  {Khalatnikov}}]{lebedev1983diagram}%
  \BibitemOpen
  \bibfield  {author} {\bibinfo {author} {\bibfnamefont {VV}~\bibnamefont
  {Lebedev}}, \bibinfo {author} {\bibfnamefont {AI}~\bibnamefont {Sukhorukov}},
  \ and\ \bibinfo {author} {\bibfnamefont {IM}~\bibnamefont {Khalatnikov}},\
  }\bibfield  {title} {\enquote {\bibinfo {title} {A diagram technique for
  hydrodynamic fluctuations},}\ }\href@noop {} {\bibfield  {journal} {\bibinfo
  {journal} {Zh. Eksp. Teor. Fiz}\ }\textbf {\bibinfo {volume} {85}},\ \bibinfo
  {pages} {1590--1601} (\bibinfo {year} {1983})}\BibitemShut {NoStop}%
\bibitem [{\citenamefont {Grozdanov}\ and\ \citenamefont
  {Polonyi}(2015)}]{Grozdanov:2013dba}%
  \BibitemOpen
  \bibfield  {author} {\bibinfo {author} {\bibfnamefont {Sa¨o}\ \bibnamefont
  {Grozdanov}}\ and\ \bibinfo {author} {\bibfnamefont {Janos}\ \bibnamefont
  {Polonyi}},\ }\bibfield  {title} {\enquote {\bibinfo {title} {{Viscosity and
  dissipative hydrodynamics from effective field theory}},}\ }\href {\doibase
  10.1103/PhysRevD.91.105031} {\bibfield  {journal} {\bibinfo  {journal} {Phys.
  Rev.}\ }\textbf {\bibinfo {volume} {D91}},\ \bibinfo {pages} {105031}
  (\bibinfo {year} {2015})},\ \Eprint {http://arxiv.org/abs/1305.3670}
  {arXiv:1305.3670 [hep-th]} \BibitemShut {NoStop}%
\bibitem [{\citenamefont {Kovtun}\ \emph {et~al.}(2014)\citenamefont {Kovtun},
  \citenamefont {Moore},\ and\ \citenamefont {Romatschke}}]{Kovtun:2014hpa}%
  \BibitemOpen
  \bibfield  {author} {\bibinfo {author} {\bibfnamefont {Pavel}\ \bibnamefont
  {Kovtun}}, \bibinfo {author} {\bibfnamefont {Guy~D.}\ \bibnamefont {Moore}},
  \ and\ \bibinfo {author} {\bibfnamefont {Paul}\ \bibnamefont {Romatschke}},\
  }\bibfield  {title} {\enquote {\bibinfo {title} {{Towards an effective action
  for relativistic dissipative hydrodynamics}},}\ }\href {\doibase
  10.1007/JHEP07(2014)123} {\bibfield  {journal} {\bibinfo  {journal} {JHEP}\
  }\textbf {\bibinfo {volume} {07}},\ \bibinfo {pages} {123} (\bibinfo {year}
  {2014})},\ \Eprint {http://arxiv.org/abs/1405.3967} {arXiv:1405.3967
  [hep-ph]} \BibitemShut {NoStop}%
\bibitem [{\citenamefont {Crossley}\ \emph {et~al.}(2017)\citenamefont
  {Crossley}, \citenamefont {Glorioso},\ and\ \citenamefont
  {Liu}}]{Crossley:2015evo}%
  \BibitemOpen
  \bibfield  {author} {\bibinfo {author} {\bibfnamefont {Michael}\ \bibnamefont
  {Crossley}}, \bibinfo {author} {\bibfnamefont {Paolo}\ \bibnamefont
  {Glorioso}}, \ and\ \bibinfo {author} {\bibfnamefont {Hong}\ \bibnamefont
  {Liu}},\ }\bibfield  {title} {\enquote {\bibinfo {title} {{Effective field
  theory of dissipative fluids}},}\ }\href {\doibase 10.1007/JHEP09(2017)095}
  {\bibfield  {journal} {\bibinfo  {journal} {JHEP}\ }\textbf {\bibinfo
  {volume} {09}},\ \bibinfo {pages} {095} (\bibinfo {year} {2017})},\ \Eprint
  {http://arxiv.org/abs/1511.03646} {arXiv:1511.03646 [hep-th]} \BibitemShut
  {NoStop}%
\bibitem [{\citenamefont {Glorioso}\ \emph {et~al.}(2017)\citenamefont
  {Glorioso}, \citenamefont {Crossley},\ and\ \citenamefont
  {Liu}}]{Glorioso:2017fpd}%
  \BibitemOpen
  \bibfield  {author} {\bibinfo {author} {\bibfnamefont {Paolo}\ \bibnamefont
  {Glorioso}}, \bibinfo {author} {\bibfnamefont {Michael}\ \bibnamefont
  {Crossley}}, \ and\ \bibinfo {author} {\bibfnamefont {Hong}\ \bibnamefont
  {Liu}},\ }\bibfield  {title} {\enquote {\bibinfo {title} {{Effective field
  theory of dissipative fluids (II): classical limit, dynamical KMS symmetry
  and entropy current}},}\ }\href {\doibase 10.1007/JHEP09(2017)096} {\bibfield
   {journal} {\bibinfo  {journal} {JHEP}\ }\textbf {\bibinfo {volume} {09}},\
  \bibinfo {pages} {096} (\bibinfo {year} {2017})},\ \Eprint
  {http://arxiv.org/abs/1701.07817} {arXiv:1701.07817 [hep-th]} \BibitemShut
  {NoStop}%
\bibitem [{\citenamefont {Stephanov}\ and\ \citenamefont
  {Yin}(2017)}]{Stephanov:2017ghc}%
  \BibitemOpen
  \bibfield  {author} {\bibinfo {author} {\bibfnamefont {M.}~\bibnamefont
  {Stephanov}}\ and\ \bibinfo {author} {\bibfnamefont {Y.}~\bibnamefont
  {Yin}},\ }\bibfield  {title} {\enquote {\bibinfo {title} {{Hydro+:
  hydrodynamics with parametric slowing down and fluctuations near the critical
  point}},}\ }\href@noop {} {\  (\bibinfo {year} {2017})},\ \Eprint
  {http://arxiv.org/abs/1712.10305} {arXiv:1712.10305 [nucl-th]} \BibitemShut
  {NoStop}%
\bibitem [{\citenamefont {Akamatsu}\ \emph {et~al.}(2017)\citenamefont
  {Akamatsu}, \citenamefont {Mazeliauskas},\ and\ \citenamefont
  {Teaney}}]{Akamatsu:2016llw}%
  \BibitemOpen
  \bibfield  {author} {\bibinfo {author} {\bibfnamefont {Yukinao}\ \bibnamefont
  {Akamatsu}}, \bibinfo {author} {\bibfnamefont {Aleksas}\ \bibnamefont
  {Mazeliauskas}}, \ and\ \bibinfo {author} {\bibfnamefont {Derek}\
  \bibnamefont {Teaney}},\ }\bibfield  {title} {\enquote {\bibinfo {title} {{A
  kinetic regime of hydrodynamic fluctuations and long time tails for a Bjorken
  expansion}},}\ }\href {\doibase 10.1103/PhysRevC.95.014909} {\bibfield
  {journal} {\bibinfo  {journal} {Phys. Rev.}\ }\textbf {\bibinfo {volume}
  {C95}},\ \bibinfo {pages} {014909} (\bibinfo {year} {2017})},\ \Eprint
  {http://arxiv.org/abs/1606.07742} {arXiv:1606.07742 [nucl-th]} \BibitemShut
  {NoStop}%
\bibitem [{\citenamefont {Kawasaki}(1970)}]{KAWASAKI19701}%
  \BibitemOpen
  \bibfield  {author} {\bibinfo {author} {\bibfnamefont {Kyozi}\ \bibnamefont
  {Kawasaki}},\ }\bibfield  {title} {\enquote {\bibinfo {title} {Kinetic
  equations and time correlation functions of critical fluctuations},}\ }\href
  {\doibase http://dx.doi.org/10.1016/0003-4916(70)90375-1} {\bibfield
  {journal} {\bibinfo  {journal} {Annals of Physics}\ }\textbf {\bibinfo
  {volume} {61}},\ \bibinfo {pages} {1 -- 56} (\bibinfo {year}
  {1970})}\BibitemShut {NoStop}%
\bibitem [{\citenamefont {Andreev}(1978)}]{andreev1978corrections}%
  \BibitemOpen
  \bibfield  {author} {\bibinfo {author} {\bibfnamefont {AF}~\bibnamefont
  {Andreev}},\ }\bibfield  {title} {\enquote {\bibinfo {title} {Corrections to
  hydrodynamics of fluids},}\ }\href@noop {} {\bibfield  {journal} {\bibinfo
  {journal} {Zhurnal Ehksperimental'noj i Teoreticheskoj Fiziki}\ }\textbf
  {\bibinfo {volume} {75}},\ \bibinfo {pages} {1132--1139} (\bibinfo {year}
  {1978})}\BibitemShut {NoStop}%
\bibitem [{\citenamefont {Akamatsu}\ \emph
  {et~al.}(2018{\natexlab{c}})\citenamefont {Akamatsu}, \citenamefont
  {Mazeliauskas},\ and\ \citenamefont {Teaney}}]{Akamatsu:2017rdu}%
  \BibitemOpen
  \bibfield  {author} {\bibinfo {author} {\bibfnamefont {Yukinao}\ \bibnamefont
  {Akamatsu}}, \bibinfo {author} {\bibfnamefont {Aleksas}\ \bibnamefont
  {Mazeliauskas}}, \ and\ \bibinfo {author} {\bibfnamefont {Derek}\
  \bibnamefont {Teaney}},\ }\bibfield  {title} {\enquote {\bibinfo {title}
  {{Bulk viscosity from hydrodynamic fluctuations with relativistic
  hydrokinetic theory}},}\ }\href {\doibase 10.1103/PhysRevC.97.024902}
  {\bibfield  {journal} {\bibinfo  {journal} {Phys. Rev.}\ }\textbf {\bibinfo
  {volume} {C97}},\ \bibinfo {pages} {024902} (\bibinfo {year}
  {2018}{\natexlab{c}})},\ \Eprint {http://arxiv.org/abs/1708.05657}
  {arXiv:1708.05657 [nucl-th]} \BibitemShut {NoStop}%
\bibitem [{\citenamefont {Ridgway}\ \emph {et~al.}(2018)\citenamefont
  {Ridgway}, \citenamefont {Rajagopal}, \citenamefont {Weller},\ and\
  \citenamefont {Yin}}]{Greg-Ryan}%
  \BibitemOpen
  \bibfield  {author} {\bibinfo {author} {\bibfnamefont {Greg}\ \bibnamefont
  {Ridgway}}, \bibinfo {author} {\bibfnamefont {Krishna}\ \bibnamefont
  {Rajagopal}}, \bibinfo {author} {\bibfnamefont {Ryan}\ \bibnamefont
  {Weller}}, \ and\ \bibinfo {author} {\bibfnamefont {Yi}~\bibnamefont {Yin}},\
  }\bibfield  {title} {\enquote {\bibinfo {title} {in preparation},}\
  }\href@noop {} {\  (\bibinfo {year} {2018})}\BibitemShut {NoStop}%
\bibitem [{\citenamefont {Lau}\ \emph {et~al.}(2018)\citenamefont {Lau},
  \citenamefont {Liu},\ and\ \citenamefont {Yin}}]{Chris}%
  \BibitemOpen
  \bibfield  {author} {\bibinfo {author} {\bibfnamefont {Chris}\ \bibnamefont
  {Lau}}, \bibinfo {author} {\bibfnamefont {Hong}\ \bibnamefont {Liu}}, \ and\
  \bibinfo {author} {\bibfnamefont {Yi}~\bibnamefont {Yin}},\ }\bibfield
  {title} {\enquote {\bibinfo {title} {{in progress}},}\ }\href@noop {} {\
  (\bibinfo {year} {2018})}\BibitemShut {NoStop}%
\bibitem [{\citenamefont {Aminov}\ \emph {et~al.}(2015)\citenamefont {Aminov},
  \citenamefont {Kafri},\ and\ \citenamefont
  {Kardar}}]{PhysRevLett.114.230602}%
  \BibitemOpen
  \bibfield  {author} {\bibinfo {author} {\bibfnamefont {Avi}\ \bibnamefont
  {Aminov}}, \bibinfo {author} {\bibfnamefont {Yariv}\ \bibnamefont {Kafri}}, \
  and\ \bibinfo {author} {\bibfnamefont {Mehran}\ \bibnamefont {Kardar}},\
  }\bibfield  {title} {\enquote {\bibinfo {title} {Fluctuation-induced forces
  in nonequilibrium diffusive dynamics},}\ }\href {\doibase
  10.1103/PhysRevLett.114.230602} {\bibfield  {journal} {\bibinfo  {journal}
  {Phys. Rev. Lett.}\ }\textbf {\bibinfo {volume} {114}},\ \bibinfo {pages}
  {230602} (\bibinfo {year} {2015})}\BibitemShut {NoStop}%
\bibitem [{\citenamefont {Carrasco}\ \emph {et~al.}(2012)\citenamefont
  {Carrasco}, \citenamefont {Hertzberg},\ and\ \citenamefont
  {Senatore}}]{Carrasco:2012cv}%
  \BibitemOpen
  \bibfield  {author} {\bibinfo {author} {\bibfnamefont {John Joseph~M.}\
  \bibnamefont {Carrasco}}, \bibinfo {author} {\bibfnamefont {Mark~P.}\
  \bibnamefont {Hertzberg}}, \ and\ \bibinfo {author} {\bibfnamefont
  {Leonardo}\ \bibnamefont {Senatore}},\ }\bibfield  {title} {\enquote
  {\bibinfo {title} {{The Effective Field Theory of Cosmological Large Scale
  Structures}},}\ }\href {\doibase 10.1007/JHEP09(2012)082} {\bibfield
  {journal} {\bibinfo  {journal} {JHEP}\ }\textbf {\bibinfo {volume} {09}},\
  \bibinfo {pages} {082} (\bibinfo {year} {2012})},\ \Eprint
  {http://arxiv.org/abs/1206.2926} {arXiv:1206.2926 [astro-ph.CO]} \BibitemShut
  {NoStop}%
\bibitem [{\citenamefont {Floerchinger}\ \emph {et~al.}(2017)\citenamefont
  {Floerchinger}, \citenamefont {Garny}, \citenamefont {Tetradis},\ and\
  \citenamefont {Wiedemann}}]{Floerchinger:2016hja}%
  \BibitemOpen
  \bibfield  {author} {\bibinfo {author} {\bibfnamefont {Stefan}\ \bibnamefont
  {Floerchinger}}, \bibinfo {author} {\bibfnamefont {Mathias}\ \bibnamefont
  {Garny}}, \bibinfo {author} {\bibfnamefont {Nikolaos}\ \bibnamefont
  {Tetradis}}, \ and\ \bibinfo {author} {\bibfnamefont {Urs~Achim}\
  \bibnamefont {Wiedemann}},\ }\bibfield  {title} {\enquote {\bibinfo {title}
  {{Renormalization-group flow of the effective action of cosmological
  large-scale structures}},}\ }\href {\doibase 10.1088/1475-7516/2017/01/048}
  {\bibfield  {journal} {\bibinfo  {journal} {JCAP}\ }\textbf {\bibinfo
  {volume} {1701}},\ \bibinfo {pages} {048} (\bibinfo {year} {2017})},\ \Eprint
  {http://arxiv.org/abs/1607.03453} {arXiv:1607.03453 [astro-ph.CO]}
  \BibitemShut {NoStop}%
\bibitem [{\citenamefont {Kharzeev}\ \emph {et~al.}(2016)\citenamefont
  {Kharzeev}, \citenamefont {Liao}, \citenamefont {Voloshin},\ and\
  \citenamefont {Wang}}]{Kharzeev:2015znc}%
  \BibitemOpen
  \bibfield  {author} {\bibinfo {author} {\bibfnamefont {D.~E.}\ \bibnamefont
  {Kharzeev}}, \bibinfo {author} {\bibfnamefont {J.}~\bibnamefont {Liao}},
  \bibinfo {author} {\bibfnamefont {S.~A.}\ \bibnamefont {Voloshin}}, \ and\
  \bibinfo {author} {\bibfnamefont {G.}~\bibnamefont {Wang}},\ }\bibfield
  {title} {\enquote {\bibinfo {title} {{Chiral magnetic and vortical effects in
  high-energy nuclear collisions?A status report}},}\ }\href {\doibase
  10.1016/j.ppnp.2016.01.001} {\bibfield  {journal} {\bibinfo  {journal} {Prog.
  Part. Nucl. Phys.}\ }\textbf {\bibinfo {volume} {88}},\ \bibinfo {pages}
  {1--28} (\bibinfo {year} {2016})},\ \Eprint {http://arxiv.org/abs/1511.04050}
  {arXiv:1511.04050 [hep-ph]} \BibitemShut {NoStop}%
\end{thebibliography}%

\end{document}